\documentclass[12pt,preprint]{aastex}
\usepackage{emulateapj5}
\newcommand{\et}{et al.}

\newcommand{\xte}{{\it RXTE}}

\newcommand{\Msun}{\hbox{$\rm\thinspace M_{\odot}$}}

\slugcomment{}
\shorttitle{X-ray Spectrum of NGC 3227}
\shortauthors{Markowitz et al.}

\begin{document}

\title{A Comprehensive X-ray Spectral Analysis of the Seyfert 1.5 NGC 3227}

\author{A. Markowitz\altaffilmark{1}, J.N. Reeves\altaffilmark{2}, I.M. George\altaffilmark{3,4}, V.\ Braito\altaffilmark{5}, R.\ Smith\altaffilmark{5}, S.\ Vaughan\altaffilmark{5}, P.\ Ar\'{e}valo\altaffilmark{6}, F.\ Tombesi\altaffilmark{4,7,8,9}
\altaffiltext{1}{Center for Astrophysics and Space Sciences, University of California, San Diego, M.C.\ 0424, La Jolla, CA, 92093-0424, USA}
\altaffiltext{2}{Astrophysics Group, School of Physical and Geographical Sciences, Keele University, Keele, Staffordshire, ST5 5BG, UK}
\altaffiltext{3}{Department of Physics, University of Maryland, Baltimore County, 1000 Hilltop Circle, Baltimore, MD, 21250, USA}
\altaffiltext{4}{X-ray Astrophysics Laboratory, Code 662, NASA Goddard Space Flight Center, Greenbelt, MD, 20771, USA}
\altaffiltext{5}{X-ray Astronomy Group, University of Leicester, Leicester LE1 7RH, UK}
\altaffiltext{6}{Department of Physics and Astronomy, University of Southampton, Highfield, Southampton, SO17 1BJ}
\altaffiltext{7}{Department of Physics and Astronomy, Johns Hopkins University, 3400 N.\ Charles Street, Baltimore, MD 21218, USA}
\altaffiltext{8}{INAF-IASF Bologna, via Gobetti 101, 40129, Bolognia, Italy}
\altaffiltext{9}{Departimento di Astronomia, Universit\`{a} degli Studi di Bologna, via Ranzini 1, 40127, Bologna, Italy} 
}

\begin{abstract}
We present results of a 100 ks {\it XMM-Newton} observation of the 
Seyfert 1.5 AGN NGC 3227.
Our best-fit broadband model to the EPIC pn spectrum consists of a moderately
flat (photon index of 1.57) hard X-ray power-law absorbed by cold 
gas with a column density of $3 \times 10^{21}$ cm$^{2}$, plus 
a strong soft excess, modeled as a steep power law with a photon index of 3.35,
absorbed by cold gas with a column density of 9 $\times$ 10$^{20}$ cm$^{-2}$. 
The soft excess increases in normalization by $\sim$20$\%$ in $\sim$20 ks, 
independently of the hard X-ray emission component, and
the UV continuum, tracked via the OM, 
also shows a strong increasing trend over the observation, consistent with reprocessing of soft X-ray emission.
Warm absorber signatures are evident in both the EPIC and RGS spectra;
we model two absorbing layers, with ionization parameters log$\xi$ = 1.2 and 2.9 erg cm s$^{-1}$, and
with similar column densities ($\sim$ 1--2 $\times 10^{21}$ cm$^{-2}$).
The outflow velocities relative to systemic of the high- and low-ionization absorbers are
estimated to be --(2060$^{+240}_{-170}$) km s$^{-1}$ and --(420$^{+430}_{-190}$) km s$^{-1}$, respectively.
The Fe K$\alpha$ line FWHM width is 7000 $\pm$ 1500 km s$^{-1}$; 
its inferred distance from the black hole is consistent with the BLR and with the inner radius of the dust
reverberation-mapped by Suganuma et al.  An emission feature near 6.0 keV is 
modeled equally well as a narrow redshifted Fe K line, possibly associated with a 
disk ``hot-spot,'' or as the red wing to a relativistically broadened Fe line profile.
{\it Swift}-BAT and archival {\it RXTE} data suggest at most a weak Compton reflection 
hump ($R \lesssim 0.5$), and a high-energy cutoff near 100 keV. From {\it RXTE} monitoring, 
we find tentative evidence for a significant fraction of the Fe line flux 
to track variations in the continuum on time scales $<$ 700 days.

\end{abstract}


\section{Introduction}

A 1993 {\it Advanced Satellite for Cosmology and Astrophysics (ASCA)} 
observation of the nucleus of the Seyfert 1.5
NGC 3227 first revealed evidence for 
ionized absorbing gas with $N_{\rm H,WA}$ $\sim$ 1--4$\times$10$^{21}$ cm$^{-2}$
(Netzer \et\ 1994, Ptak \et\ 1994). Komossa \& Fink (1997b),
using 1993 {\it R\"{o}ntgen Satellite (ROSAT)}-PSPC data, and George \et\ (1998b), 
using {\it ASCA} data taken in 1993 and 1995,
further modeled the warm absorber, finding it to be relatively
lowly-ionized relative to most other Seyferts' warm absorbers. 
Given the low-resolution spectra,
the solution for further progress was to obtain a gratings spectrum.
Komossa (2002) reported on a {\it Chandra}-LETGS observation of NGC 3227
in October 2000, noting edge-like features near 0.7 keV. 
{\it XMM-Newton} first observed NGC 3227 in 
November 2000, when the source was undergoing a 3-month long period
of very high levels of obscuration due to a compact cloud with 
$N_{\rm H}$ near $3 \times 10^{23}$ cm$^{-2}$, covering 90$\%$ of the central source, and
likely associated with the BLR, traversing the line of sight (Lamer \et\ 2003). 
The soft X-ray data from this observation indicated the 
warm absorber to have ionization parameter log$\xi$\footnote{The ionization parameter $\xi \equiv L_{\rm ion} n_{\rm e}^{-1} r^{-2}$, where $L_{\rm ion}$ is the isotropic 1--1000 Ryd ionizing continuum luminosity, $n{\rm e}$ is the electron number density, and $r$ is the distance from the central continuum source to the absorbing gas.} = 1.7--2.0 erg cm s$^{-1}$ and
a column density $N_{\rm H,WA}$ = 2--9$\times$10$^{21}$ cm$^{-2}$ (Gondoin \et\ 2003).

Optical spectra of the nucleus of NGC 3227 have indicated strong reddening
due to the presence of dust along the line of sight.
The narrow line H$\alpha$/H$\beta$ ratio, for instance, indicates reddening,
with measured ratios in NGC 3227 implying visual extinctions $A_{\rm V}$ =  (H$\alpha$/H$\beta$)/3.1 
near 1.2--1.7 (Cohen 1983; Gonzalez Delgado \& Perez 1997),
though $A_{\rm V}$ values of 4.5--4.9 have been reported
(Mundell \et\ 1995a, Rubin \& Ford 1968)\footnote{The 
H$\alpha$/H$\beta$ ratio is intrinsically 3.1, assuming case B recombination and assuming that collisional excitation 
is negligible, applicable for typical NLR conditions, but see warnings by Netzer 1990 regarding
uncertainties in the intrinsic Balmer decrement.}.
The broad H$\alpha$/H$\beta$ line ratio shows a similar level of reddening,
$A_{\rm V}$ = 1.4 (Cohen 1983, Winge \et\ 1995).
The steep drop in NGC 3227's continuum flux from the near-UV 
to the far-UV also supports reddening: the spectral index from {\it International Ultraviolet Explorer (IUE)}
measurements is $\alpha_{\rm IUE} = -2.9$ for NGC 3227, whereas $\alpha_{\rm IUE} = -1.4$
is more typical for
Seyfert 1s (Komossa \& Fink 1997b; Courvoisier \& Paltani 1992; Kinney \et\ 1991).
Assuming a standard Galactic gas/dust ratio of
$N_{\rm H}$/$A_{\rm V}$ = 2$\times$10$^{21}$ cm$^{-2}$ magn$^{-1}$
(e.g., Shull \& van Steenburg 1985), the amount of NLR and BLR
reddening implies that the dust is associated with 
a column of gas with $N_{\rm H}$ $\sim$ 2--3 $\times$10$^{21}$ cm$^{-2}$.

Komossa \& Fink (1997b) and George \et\ (1998b) also reported local absorption due to cold material
with a column density $N_{\rm H}$ = 3--6 $\times$10$^{20}$ cm$^{-2}$,
in addition to the Galactic column (2.1$\times$10$^{20}$ cm$^{-2}$,
Dickey \& Lockman 1990, Murphy \et\ 1996). The inferred local cold column density
agrees with that derived from 21cm VLA observations (Mundell \et\ 1995b);
however, it is not sufficient to produce the observed optical reddening,
assuming a standard Galactic gas/dust ratio.
The agreement between the inferred column density associated with the
dust and the column density of the ionized gas, however, prompted Komossa \& Fink (1997b) to 
suggest that the dust was embedded in the warm absorber and not
associated with the cold gas. Brandt, Fabian, \& Pounds (1996) first suggested 
this "dusty warm absorber" concept for the quasar IRAS 13349+2438.


However, Kraemer \et\ (2000) suggested that it was
unlikely dust could survive within the high-ionization gas.
They proposed a configuration consisting of {\it two}
warm absorber zones, in addition to the cold column.
One warm absorber is dust-free and highly-ionized,
responsible for the oxygen edges; the other is a very lowly-ionized absorber with dust mixed in,
referred to as a "dusty lukewarm absorber." Those authors 
modeled the latter absorber with a density of 20 cm$^{-3}$ and 
a location encompassing or lying just outside the NLR, which has a 
radius of $\sim$100 pc (based on [O III] imaging; Schmitt \& Kinney 1996).
Using {\it Hubble Space Telescope (HST)}-STIS, Crenshaw \et\ (2001) first observed
the features of the dusty lukewarm absorber, evident in the form of 
optical/UV absorption lines due to intermediate species of C, N and Si;
Crenshaw \& Kraemer (2001)
suggested that the absorber may possibly be a relatively highly-ionized
component of the NLR seen in absorption. 
Kraemer \et\ (2000) suggested that the dust may evaporate off the 
putative molecular torus and be swept up in an outflowing wind
(e.g., Reynolds 1997).
Based on V-band-to-K-band reverberation mapping (the latter band dominated by
thermal dust emission) conducted by the MAGNUM collaboration, 
Suganuma \et\ (2006) concluded that 
the innermost extent the dust was at a radius of $\sim$5--20 light-days,
suggesting an alternate or additional site from which an outflowing wind 
could sweep up dust.

In 2006, we obtained a new, 100 ks {\it XMM-Newton} observation of NGC 3227 in an unobscured
state, to better characterize the broadband X-ray continuum and X-ray
warm absorber features, search for signatures of dust in the X-ray spectrum, 
relate the X-ray and dusty optical/UV warm absorbers,
and study Fe K bandpass emission features.
$\S$2 describes the observations and data reduction.
$\S$3 and $\S$4 describe EPIC spectral fits to the time-averaged spectrum, 
focusing on the Fe K bandpass and then extending to the 
0.2--10 keV bandpass, respectively.
$\S$5 describes the fits to the RGS spectra. 
$\S$6 describes spectral analysis of archival {\it RXTE} data to constrain
the hard X-ray continuum.
$\S$7 describes time- and flux-resolved spectral fits to the EPIC data,
including $F_{\rm var}$ spectra, and time-resolved spectral fits
to the {\it RXTE} data to quantify the temporal behavior of the Fe line.
X-ray/UV correlations are presented in $\S$8.
The results are discussed in
$\S$9, and a brief summary is given in $\S$10.

\section{Observations and Data Reduction}

\subsection{{\it XMM-Newton} data reduction}

NGC 3227 was observed by {\it XMM-Newton} during Revolution 1279, on 
December 3--4, 2006, for a duration of 99 ks.
{\it XMM-Newton}'s European Photon Imaging Camera (EPIC) consists of one pn CCD
back-illuminated array sensitive to 0.15--15 keV photons (Str\"{u}der
\et\ 2001), and two MOS CCD front-illuminated arrays sensitive to
0.15--12 keV photons (MOS 1 and MOS 2, Turner \et\ 2001).
Data from the pn were taken in Large Window mode; data from both MOSes
were taken in PrimePartialW2/small window mode. 
The medium filter was used for all three EPIC instruments.
Spectra were extracted using {\it XMM-Newton}-SAS version 7.1.0 and using standard
extraction procedures. For all three cameras, source data were
extracted from a circular region of radius 40$\arcsec$; backgrounds were
extracted from circles of identical size, centered $\sim$3$\arcmin$ away
from the core. Hot, flickering, or bad pixels were excluded. Data were
selected using event patterns 0--4 for the pn and 0--12 for the MOSes. 
Using the SAS task {\sc epatplot}, we verified that pile-up was negligible
for the pn. For the MOS, pile-up was minimal ($<$3$\%$) up to 8 keV,
and a bit higher (3--10$\%$) above 8 keV.
The effect of such pile-up is to artifically flatten the spectrum.
However, we also extracted counts from the MOS CCDs
using an annular region with inner radius 5$\arcsec$ and outer
radius 40$\arcsec$ to minimize pile-up; the resulting spectra
were virtually identical to those extracted using the circular region.
We elected to fit the spectra extracted using circular regions
in order to maximize the photon statistics associated with 
emission features in the Fe K bandpass.

We inspected the 10--12 keV pn background
light curve; there were no significant flares (the 10--12 keV rate
never exceeded 0.04 ct s$^{-1}$). The total good exposure time 
was 90.7 ks for pn and 96.4 ks for each MOS.
The 0.2--10 keV pn light curve, binned to 600 s and normalized by its
mean count rate (11.74 ct s$^{-1}$), is displayed in the top
panel of Figure 1. Similarly displayed in Figure 1 are the 0.2--1 and 3--10 keV
light curves; in these energy ranges, the total continuum is dominated
by the soft excess and hard X-ray power-law components, respectively,
as will be demonstrated in $\S$4.

The Reflection Grating Spectrometer (RGS) data were extracted using standard extraction
procedures, including the SAS task {\sc rgsproc} and the most recent calibration files.
We used the first-order data only; spectra were binned every 10 channels ($\sim$0.1$\AA$, or roughly the RGS resolution).
Virtually all bins contained at least 20 counts, allowing use of the $\chi^2$ statistic.
Bad RGS channels were ignored. The good exposure time was 99.2 ks for each RGS.

The Optical Monitor (OM) was in Science User Defined
mode, with one ``image'' window $5.17\arcmin \times 5.17\arcmin$ centered on NGC 3227.
The UVW1 filter, whose effective area peaks near 260 nm, was used throughout.
Extraction proceeded in a manner similar to $\S$2--3 of Smith \& Vaughan (2007). 
Source photons were collected using a 12-pixel radius circle centered on NGC 3227.
Emission from the host galaxy was not subtracted; the effect of including
such a constant component is that any variability we observe is thus 
likely a lower limit to the intrinsic level of variability associated with the AGN. 
The background was extracted from a circular region away from the host galaxy.
The resulting net light curve, extracted in 1400 s bins,
is displayed in Figure 1.

\subsection{{\it RXTE} PCA and HEXTE extraction}

To investigate the form of the $>$10 keV continuum and estimate
the strength of any Compton reflection hump present, we
examined archival 
Rossi X-ray Timing Explorer {\it RXTE} Proportional Counter Array (PCA;
Swank \et\ 1998)
and High-Energy X-Ray Timing Experiment (HEXTE; Rothschild \et\ 1998) monitoring data.
NGC 3227 was observed in November 1996 for $\sim$260 ks.
NGC 3227 was also monitored from 1999 January 2 until 2005 December 4,
with one visit every 2--7 days, along with a period of more intensive monitoring 
approximately four times daily from 2000 April 2 to 2000 June 7
(see Uttley \& M$^{\rm c}$Hardy 2005 for further details regarding sampling patterns).
Each monitoring snapshot typically lasted 1 ks.
There were no observations simultaneous with the 2006 {\it XMM-Newton} observation, so we considered
all available data in the public archive in order to estimate average long-term properties.
However, data taken from late 2000 to early 2001 (modified Julian day [MJD] 51850--52050), 
affected by the passage of the compact cloud across the line of sight, were ignored
during spectral fitting.

PCA data were extracted and screened using standard 
methods and tools. 
The 'L7-240' background models, appropriate for faint sources, were used.
See e.g., Markowitz, Edelson \& Vaughan (2003) 
for details on PCA data extraction and background subtraction, the 
dominant source of systematic uncertainty (e.g., in total broadband count rate)
in these data.
Counts were extracted from Proportional Counter Units (PCUs) 0--2, 0 and 2, and 2 only for data
observed before 1999 December 23, between 1999 December 23 and 2000 May 12, and after
2000 May 12, respectively. 
Response files were generated for gain epochs 3, 4 and 5 separately
(data observed before 1999 March 22, between 1999 March 22 and 2000 May 12, and
after 2000 May 12, respectively).
As the response of the PCA slowly hardens
over time due to the gradual leak of propane into the xenon layers,
data within each gain epoch were further split into roughly equal segments, each
spanning durations of roughly
a couple hundreds days, and separate response files were
generated for each segment. Spectra and responses were then added,
weighting by exposure times, using the FTOOLS {\sc sumpha} and {\sc addrmf}, respectively.
The total exposure time for all PCA data was 842.7 ks.

The HEXTE instrument consists of
two independent clusters (A and B), each containing four scintillation 
counters  which share a common 1$\degr$ FWHM 
field of view. Source and background spectra were extracted from each 
individual \xte\ visit using Science Event data and standard extraction 
procedures. The same good time intervals used for the PCA data (e.g., 
including Earth elevation and SAA passage screening) were applied to the 
HEXTE data. To measure real-time background measurements, the two HEXTE 
clusters each undergo two-sided rocking to offset positions, in this 
case, to 1.5$\arcdeg$ off-source, switching every 32 seconds
(16 seconds before 1998 Jan.). No strong 
contaminating hard X-ray source within 2$\degr$ is evident
(e.g., from the {\it RXTE}-PCA or {\it Swift}-BAT all-sky slew survey data
available at the HEASARC's online SkyView service).
Cluster A data taken between 
2004 Dec 13 and 2005 Jan 14 were excluded, as the cluster did not rock 
on/off source. Detector 2 aboard Cluster B lost spectral capabilities
in 1996; these data were excluded from spectral analysis. Cluster A and B 
data were extracted separately and not combined, as their response matrices 
differ slightly.  Deadtime corrections were applied.
All data within each cluster were combined,
except for the 1996 data; 
that is, the 1996 (16 s rocking)
and 1999--2005 (32 s rocking) spectra were fit separately.
We note that the 1996 and 1999--2005 spectra agree well with each other for each cluster,
illustrating HEXTE's performance down to source fluxes which are
1$\%$ of the background.
Good exposure times for the four HEXTE source spectra
were 237.1 ks (cluster A) and 234.2 ks (cluster B)
for the 1999--2005 data, and 41.5 (A) and 41.8 ks (B) for
the 1996 data. All data were grouped as follows:
channels 17--30, 31--39, 40--47, 48--67, 68--79 and $\geq80$
were grouped by factors of 2,3,4,5,6 and 10, respectively.
The LLAGN/LINER source NGC 3226 is located about 2$\arcmin$ away and is thus
in the field of view of both the PCA and HEXTE, but its flux
is a factor of $\sim$ 50 fainter than that of NGC 3227 in the 2-10 keV band
(Binder et al, in prep.; George \et\ 2001).

\section{EPIC pn spectral fits to the Fe K Bandpass}

{\sc xspec} (Arnaud 1996) v.11.3.2ag was used for all spectral fitting.
All errors below correspond to $\Delta$$\chi^2$ = 2.71
(90$\%$ confidence level for one interesting parameter when the errors are symmetric) unless
otherwise noted.
The abundances of Lodders (2003) were used.
NGC 3227's redshift is 0.00386 (de Vaucouleurs et al.\ 1991);
a distance of 20.3 Mpc was used (Mould et al.\ 2000)
was used to calculate luminosities.
In all models, we included a
column of neutral absorption fixed at the Galactic column
of 2.1$\times$10$^{20}$ cm$^{-2}$ (Dickey \& Lockman 1990, Murphy \et\ 1996). 

We started by fitting the EPIC-pn data, 4--10 keV only.
Residuals to a simple power-law fit
($\chi^2$/$dof$ = 1515/999) showed an obvious Fe K$\alpha$ emission
line, but also some interesting structure near 6.0 keV; see Figure 2a.
In our fits, we tested two competing models that explain the
emission features roughly equally well. As explained below,
in both models, narrow emission features are detected at
6.40 keV and 7.37 keV (Fe K$\alpha$ and Ni K$\alpha$ respectively).
The models differ in how emission near 6.0 keV is modeled:
as a narrow Gaussian emission line, or as a relativistic diskline
(hereafter ``GA'' and ``DL'' models). 

We started with the former case. First, we included
a Gaussian component at 6.40 keV to model the
Fe K$\alpha$ line. In the best-fit ``GA'' model,
the Fe K$\alpha$ line best-fit energy, intensity and $EW$
were 6.403$^{+0.011}_{-0.010}$ keV, consistent with neutral Fe. 
Its intensity and $EW$ were 
3.5$\pm$0.4 $\times 10^{-5}$ ph cm$^{-2}$ s$^{-1}$ and 91$\pm$10 eV,
respectively. Its width $\sigma$ was 65$\pm$14 eV, which corresponds to a
FWHM velocity of 7000$\pm$1500 km s$^{-1}$.

Figure 3 shows contour plots resulting from adding a 
``sliding Gaussian'' to the ``power-law + Fe K$\alpha$ line'' model 
to trace the emission residuals after the narrow Fe K$\alpha$ line
was modeled. That is, we added a Gaussian with width $\sigma$ fixed at 10 eV
and searched over 4--9 keV in units of 0.1 keV, i.e., in
steps smaller than the instrumental resolution. The narrow feature
near 6.0 keV is clear; other residuals near 7.0 and 7.4 keV
also warrant further investigation.

To model Fe K$\beta$ emission,
we added a Gaussian to the model, keeping the   
energy centroid fixed and width $\sigma$ tied to that of the Fe K$\alpha$ line.
However, the line was not significantly detected according to an $F$-test.
The upper limit to the line's $EW$ was 15 eV;
the upper limit on the K$\beta$/K$\alpha$ normalization ratio
was 0.19 (yielding no useful constraints on the ionization 
state of the Fe-line emitting material).  
We included this feature in the rest of the fits for completeness,
with intensity fixed to 0.13 times that of the Fe K$\alpha$ line.
Residuals to a model consisting of the power-law and
the K$\alpha$ and K$\beta$ lines (where $\chi^2$/$dof$ was 1028.3/996)
are plotted in Figure~2b.

We next added an Fe K edge at 7.11 keV (energy fixed) 
likely associated with reflection; $\chi^2$/$dof$ dropped to 1017.6/995; 
residuals to a model including the edge are plotted in Figure~2c.
In the best-fit model, the optical depth $\tau$ was 0.05$\pm$0.03.

Next, we used a Gaussian component
to model an emission line at 7.39$^{+0.08}_{-0.07}$ keV,
with width $\sigma$ tied to that of the Fe K$\alpha$ line,
consistent with emission from Ni K$\alpha$.
Its intensity and equivalent width were 
$4^{+4}_{-3} \times 10^{-6}$ ph cm$^{-2}$ s$^{-1}$ and 
$13^{+13}_{-10}$ eV, respectively.  $\chi^2$/$dof$ dropped to 1010.2/993;
it was significant at 97.3$\%$ 
confidence according to an $F$-test to include this line.
Residuals to a model including this line are plotted in Figure~2d.

Next, we added a narrow Gaussian emission profile near 6.0 keV 
to model those residuals;
$\Delta$$\chi^2$ was --24.94 for 3 less $dof$,
significant at $>$99.99$\%$ confidence according a standard $F$-test.
This line is not resolved 
(width $\sigma$ $<$ 200 eV). Its intensity and 
$EW$ are $9^{+8}_{-4} \times 10^{-6}$ ph cm $^{-2}$ s$^{-1}$ and     
21$^{+19}_{-9}$ eV, respectively.   
A contour plot of intensity versus energy (confidence levels are for
two interesting parameters) is shown in Figure 4.
The best-fit line energy is 6.04$^{+0.18}_{-0.04}$ keV
inconsistent with 6.24 keV, the lowest photon energy 
associated with a Compton shoulder scattering feature.
Using a Gaussian to model a Compton shoulder, we should expect 
a centroid energy $>$ 6.24 keV; fixing the Gaussian line energy at 6.24 keV
failed to fully model away the 6.0 keV residuals.
We performed Monte Carlo simulations to assess the statistical
significance of this emission feature, as standard usage of 
the $F$-test may overestimate the statistical significance of
lines at arbitrary energies such as this one (see e.g., warnings by Protassov \et\ 2002).
The simulations were conducted following $\S$4.3.3 of Markowitz, Reeves \& Braito (2006); see
also $\S$3.3 of Porquet \et\ (2004). 
The simulations indicated that the likelihood that
the 6.0 keV emission line is photon noise is $<$0.1$\%$, i.e., the line is
significant at $>$99.9$\%$ confidence.

In this best-fit ``GA'' model, $\chi^2$/$dof$ was 983.2/990;
$\Gamma$ was 1.57$\pm$0.03. Other best-fit parameters are shown in Table~1.
Residuals are plotted in Figure~2e.

Next, we tried to model the 6.0 keV feature as the red wing
of a relativistic diskline profile. 
We used a {\sc Laor} model profile (Laor 1991) to 
model the diskline, with emissivity index 
$\beta$\footnote{radial emissivity per unit area is quantified as a power law, $r^{-\beta}$}
fixed at 3. $\chi^2$/$dof$ was 988.7/989 ($\Delta$$\chi^2$ = --16.8;
significant at 99.8$\%$ confidence in an $F$-test)
in the best-fit ``DL'' model.  The $EW$ of the diskline was 81$^{+42}_{-30}$ eV;
its inclination was $<$25$\degr$, and the inner radius was $<$22 $R_{\rm g}$\footnote{1 $R_{\rm g} \equiv GM_{\rm BH}/c^2$}. Visually, there still remained correlated, emission-like residuals from 5.9--6.1 keV,
as shown in Figure 2f, which might at first 
suggest that the 6.0 keV emission feature may be better modeled
as a narrow feature than as a red wing of a diskline. However, given the similar values of
$\chi^2$/$dof$ and the instrument resolution, it would be difficult to demonstrate
that the difference in residuals between the two models is not consistent with 
photon noise. 


\subsection{MOS 1+2 spectral fits to the Fe K bandpass} 

As a double-check on spectral modeling, we applied
our best-fit ``GA'' model to both time-averaged
MOS spectra (fit simultaneously, with all line 
parameters tied between both MOS spectra). The power-law 
indices and normalizations were allowed to differ between MOS 1
and MOS 2. Overall,
the MOS and PN yielded qualitatively similar results. 
A simple power-law fit resulted in $\chi^2$/$dof$ = 771.9/558; 
data/model residuals are plotted in the top panel of Figure 5.
Adding Gaussian components to represent Fe K$\alpha$ and K$\beta$ emission
in a manner similar to the pn fits caused $\chi^2$/$dof$ to drop to
546.86/555. Again we performed a ``sliding narrow ($\sigma$ = 10 eV)
Gaussian'' test; the resulting contours are plotted in Figure 6 and again indicate
emission near 6.0-6.1 keV. 

Finally, adding a narrow Gaussian (width $\sigma$ fixed
at 10 eV) at $6.11^{+0.06}_{-0.13}$ 
keV further improved the fit; $\chi^2$/$dof$ fell to 
535.87/553 ($\Delta\chi^2 = -11.0$ for two less $dof$; 
significant at 99.7$\%$ confidence
according to an $F$-test used in the standard way). 
The best-fit values of the intensity and $EW$ were $6.4 \pm 3.4 \times 10^{-6}$ ph cm$^{-2}$ s$^{-1}$
and $13 \pm 7$ eV, respectively.
Data/model residuals for the best-fit model are plotted in the bottom panel of
Figure 5 (There remain $\sim$10--15$\%$ 
residuals in ratio space near 7.0 and 7.5 keV
in Figure 5, but Figure 6 indicates that those features are not detected
with high significance.). 
A contour plot of the intensity of the emission line at 6.11 keV versus line energy
is plotted in Figure 7.
Again we performed Monte Carlo simulations
to gauge the detection significance, finding the line to be detected at 95$\%$ confidence
in the MOS 1+2 spectrum i.e., independent of the pn detection. 
In fact, we can multiply the independent null hypothesis probabilities
for the pn and MOS ($<$ 0.001 and 0.05) 
to yield a ``effective'' null hypothesis probability of $<$ $5\times10^{-5}$.

The detection of features at 6.0--6.1 keV in simultaneous pn and
MOS 1+2 spectra illustrates the importance of observing with 
multiple X-ray instruments to attempt to
distinguish between narrow and/or weak features which may be intrinsic to the target and 
features which are artifacts associated with photon noise
(even when Monte Carlo simulations are used to gauge the detection
significances in individual spectra). The likelihood that a strong emission feature could appear 
at the same energy in both the pn and the MOS 1+2 spectra and be due to photon noise
is likely quite small.  

The photon indices for the MOS 1 and MOS 2  
spectra were 1.37$\pm$0.04  and 1.46$\pm$0.04, respectively; the flatter
values of the photon index compared to the pn are
likely the result of pile-up (MOS spectra extracted using only pattern 0 events
has photon indices which were $\sim$0.3 steeper than the MOS pattern 0--12 spectra).
Finally, we added an Fe K edge (energy fixed at 7.11 keV);
$\chi^2$/$dof$ fell to 517.4/552 for $\tau = 0.09 \pm 0.04$.
Emission from Ni K$\alpha$ was not significantly
detected. All other parameters were consistent at 
the 90$\%$ confidence level with those measured by the pn.

We also applied the best-fit ``DL'' model to the MOS 1+2 spectrum.
$\chi^2$/$dof$ was 504.9/550, with all diskline and
Fe K$\alpha$ line parameters consistent with those in the pn fits.


\section{EPIC pn Broadband Spectral Modeling}

We started with the best-fit ``GA'' model and
included data down to 0.2 keV. 
In addition to the Galactic column, we included
a {\sc zwabs} component at the systemic redshift
to model any excess cold absorption 
associated with e.g., the host galaxy or circumnuclear material,
$N_{\rm H,local}$.

The residuals (Figure 8a) showed a large soft excess 
and a large absorption trough at 
$\sim$0.73 to $\sim$0.92 keV which we identify as 
an Fe UTA feature. A narrow absorption feature near 
1.34 keV is apparent, likely due to Mg XI.
A narrow absorption feature near 1.85 keV could be due to Si XIII, 
but could also be due to calibration uncertainties associated with 
the instrumental Si K edge.

We first modeled the soft excess using a steep power-law with
photon index $\Gamma_{\rm SX}$ near 3. Our final, best-fit model (including
modeling the warm absorbers and \ion{O}{7} emission line; see below) 
is henceforth referred to as the ``SXPL (soft X-ray power law)'' model.
In the best-fit SXPL model, $\Gamma_{\rm SX}$ was 3.35$^{+0.27}_{-0.10}$.
We also added another layer of cold 
absorption at the systemic redshift, applying it only to the hard X-ray power 
law; in the best-fit SXPL model, its column density $N_{\rm H,HX}$ was 
2.9$^{+0.3}_{-0.8}$$\times$10$^{21}$ cm$^{-2}$.
This improved the overall fit substantially 
($\chi^2$/$dof$ fell to 3070.3/1730), but the 0.7--0.9 keV trough remained; see
Figure 8b. In this model, $\Gamma_{\rm HX}$ was 1.64;
we note that forcing the photon indices of
the soft and hard power-law components $\Gamma_{\rm SX}$
and $\Gamma_{\rm HX}$ to be equal resulted in
a very poor fit, with $\chi$/$dof$ =  6057/1731.

Next, we used an {\sc xstar} grid, which 
assumed a turbulent velocity width of 100 km s$^{-1}$,
to model a layer of warm absorption with ionization 
parameter log$\xi$ and column density $N_{\rm H}$.
The grid assumed an underlying optical to X-ray spectral energy distribution described as follows:
Below 0.001 keV, a power-law component with spectral index $\alpha$ = 1.0 ($\Gamma=2.0$);
from 0.001 to 0.04 keV, a power law with $\alpha$ = 0.2 ($\Gamma=1.2$), following e.g.,
Elvis \et\ (1994) and Netzer (1996);
from 0.04 to 1.0 keV, a steep power law with $\alpha$ = 1.9 ($\Gamma=2.9$; the intrinsic EUV continuum is not well studied,
but we assume that the soft excess seen in the {\it XMM-Newton} spectrum extends down to 0.04 keV);
and above 1.0 keV, a power law with $\alpha$ = 0.5 ($\Gamma=1.5)$, with the
$>$0.04 keV power law indices based on the best-fitting SXPL model.
Initially, solar abundances and a zero velocity offset relative to systemic
were assumed. Applying one zone of warm absorption to the model,
$\chi^2$/$dof$ fell to 2184.1/1728 for log$\xi$ $\sim$ 2.2 and
$N_{\rm H}$ $\sim$ 2$\times$10$^{21}$ cm$^{-2}$. As shown in Figure 8c, however,
significant data/model residuals remained. A model incorporating
two zones of warm absorption successfuly modeled all the absorption-like residuals,
including the trough near 0.75 keV; $\chi^2$/$dof$ fell to 1942.1/1726.
In this model, the two X-ray absorbers had ionization parameters
log$\xi_{\rm lo}$ near 0.3 and log$\xi_{\rm hi}$ near 2.5.
We henceforth refer to the two X-ray absorbers as the
low-ionization X-ray absorber (though the ionization parameter
is still higher than that of the optical/UV lukewarm absorber; see $\S$9),
and high-ionization X-ray absorber, respectively.

By now, the residuals (see Figure 8d) showed a narrow emission 
feature near 0.56 keV, likely due to \ion{O}{7}. 
Adding a narrow Gaussian component 
at 0.58$\pm$0.01 keV with width $\sigma$ fixed at 1 eV caused $\chi^2$ to drop
by 75 for two less $dof$. 
However, we note that the warm absorbers each predict 
narrow absorption lines due to \ion{O}{7}, so there are likely large systematic 
uncertainties associated with the measured intensity of the \ion{O}{7} emission line.
This model, with $\chi^2$/$dof$=1866.9/1724,
is our best-fit  SXPL  model. In this model,
the column densities and ionization parameters of the 
low- and high-ionization absorbers are:                                        
$N_{\rm H,lo} = 1.0^{+0.3}_{-0.1} \times 10^{21}$ cm$^{-2}$,
log$\xi_{\rm lo}$ = $1.45^{+0.16}_{-0.07}$,
$N_{\rm H,hi} = 1.8^{+1.2}_{-0.6} \times$10$^{21}$ cm$^{-2}$,
and log$\xi_{\rm hi}$ = $2.93^{+0.15}_{-0.09}$.
Other best-fit parameters are listed in Table 2.
The data/model residuals are shown in Figure 8e.
Residuals near 1.8 and 2.2  keV are instrumental
(Si K and Au M). Narrow emission-like features near 0.42 and 0.92 
are tempting to identify as \ion{N}{6} and \ion{Ne}{9} emission, but they are
likely narrower than the instrument resolution. 
There also appear to be some negative residuals near 6.8 keV, close to
the energies associated with \ion{Fe}{25} and \ion{Fe}{26}. However,
fitting this feature with an inverted Gaussian does not yield
significant improvement to the fit; it is likely a spurious feature
and will not be discussed further.

Note that residuals near the expected O and Fe L3
edge energies due to dust (0.53 and 0.71 keV) are fine; we do not require
additional edges at those energies (though we revisit this issue with the
higher resolution RGS data below).
If dust is present in the warm absorber,
abundances of O and Fe (and also Si and C)
may be expected to be low due to depletion onto grains (see Snow \& Witt 1996).
However, the measured O and Fe abundances of the warm absorber
are consistent with solar, and left fixed at solar values for
the remainder of the paper.

In the best-fit SXPL model, the hard X-ray power-law component has
a photon index $\Gamma_{\rm HX} = 1.57 \pm 0.02$,
a couple tenths lower than ``canonical'' values of 1.8--1.9 in many broad-line
Seyferts. We therefore explored the possibility that the observed
low value of the photon index is an artifact caused by a 
yet-unmodeled partial-covering absorbing component
with column density $N_{\rm H,PC} \sim 10^{23}$ cm$^{-2}$ 
along the line of sight to the nucleus.
Such material is expected to exist within light-days
of the black hole, as demonstrated by the 2000-1 obscuration event.
Furthermore, such a column is required to explain the origin of the 
Fe K$\alpha$ line (see $\S$9.4) in the absence of strong Compton 
reflection (see $\S$6).
Modifying the hard X-ray power-law component
to be absorbed by a partial-covering neutral
absorber with $N_{\rm H,PC}$ 
set to 0.3, 1.0 or 3.0 $\times 10^{23}$ cm$^{-2}$, we find 
only upper limits to the covering fraction in the range 5--15$\%$. 
In the best fit models, $\Gamma_{\rm HX}$ was never higher than about 1.6.
Partial-covering models are not analyzed further.

As before, we applied our best-fit SXPL model to the time-averaged
MOS 1+2 spectra as a double-check. 
The MOS fits, using this model and fitting over 0.4--10 keV,
and allowing power-law component parameters to differ between MOS 1 and MOS 2, 
yielded $\chi^2$/$dof$ = 1406.33/1017, and
required slightly higher values of the column density and ionization parameter
of the high-ionization warm absorber 
($N_{\rm H,hi} = 1.2^{+0.3}_{-0.5} \times 10^{21}$ cm$^{-2}$ and 
log$\xi_{\rm hi}$ = 3.22$^{+0.04}_{-0.11}$ erg cm s$^{-1}$, respectively)  
compared to the pn.
$\Gamma_{\rm SX}$ for MOS1 and MOS2 were consistent with
that for the pn. $\Gamma_{\rm HX}$ for MOS1 was $1.52 \pm 0.01$, slightly flatter
than that for the pn. $\Gamma_{\rm HX}$ for MOS2 was $1.57 \pm 0.02$, consistent with the pn.
All other parameters were consistent at 
the 90$\%$ confidence level with those measured by the pn.

\subsection{Alternate parametrizations of the soft excess}

We returned to the pn data to consider alternate methods of modeling
the soft excess. An alternate parametrization of the soft excess, a blackbody,
yields a nearly identical value of $\chi^2$/$dof$ (1870.4/1724).
Data/model residuals are plotted in Figure 8f.
The blackbody temperature $k_{\rm B}T$ = 83$^{+1}_{-4}$ eV 
in the pn fit ($77 \pm 3$ eV for the MOS), similar            
to values found for model fits
incorporating blackbody components for most other Seyferts,
e.g., $k_{\rm B}T$ $\sim$100--150 eV for many narrow-line Seyfert 
1s. It is generally accepted now that a blackbody is a physically unplausible
description of the soft excess in AGN.
Bechtold et al.\ (1987) were the first to point out 
that the best-fit blackbody temperatures are generally too high to
be associated with accretion disks around supermassive black holes.
The consistency of blackbody temperatures across a wide range of 
Seyfert 1 properties, including black hole mass, 
furthers suggests that a blackbody is an unphysical
parametrization (e.g., Gierli\'{n}ski \& Done 2004).
However, the best-fit blackbody temperature in the current fit is
consistent with previous results for NGC 3227 (e.g., Komossa \& Fink 1997b).
We note that in this model (henceforth denoted simply as the
``BB'' model), $N_{\rm H,HX}$ has dropped to 
$\leq$ $N_{\rm H,local}$,
i.e., the value of $N_{\rm H,HX}$ is model dependent.

We next tried fitting the soft excess using
the Comptonization model {\sc CompST} (Sunyaev \&
Titarchuk 1980). Such components have been used 
e.g., by Gierli\'{n}ski \& Done (2004) to
model the soft excess emission as Comptonization of
accretion disk seed photons in
a cool ($k_{\rm B}T \sim 0.3$ keV), 
optically-thick corona, distinct from the optically-thin, hot 
($k_{\rm B}T \sim 100$ keV) corona
generally thought to be responsible
for the hard X-ray power-law component in Seyferts.
Physically, such a component could be potentially 
associated with the transition region between 
an optically-thick accretion disc and 
an optically-thin inner corona 
(Magdziarz et al.\ 1998), or in the hot, ionized surface of
the accretion disk (Hubeny \et\ 2001; Janiuk, Czerny \& Madejski 2001).
However, similar to the blackbody component, using such Comptonization
components to model soft excesses tends to yield a 
narrow range of best-fit temperatures across a range
of Seyfert properties.
In our best-fit model (``COMPST'', where $\chi^2$/$dof$ = 1863.4/1723), 
$k_{\rm B}T = 0.35^{+0.02}_{-0.03}$ keV,
and optical depth $\tau = 24^{+2}_{-4}$,     
similar to previous fits to other AGN
(however, see e.g., Vaughan \et\ 2002 for warnings regarding
the covariance of optical depth and temperature in this model).
Best-fit parameters for the models using
the blackbody and {\sc CompST} components are listed in Table 2,
though the reader must bear in mind the difficulties
each of these models faces in terms of physical plausibility.
Data/model residuals are not plotted, as they are virtually identical
to those for the SXPL model (Figure 8e).

Figure 9 shows unfolded model spectra for the SXPL, BB and COMPST            
models, demonstrating how the steep soft excess is produced in each
(and demonstrating why the value of $N_{\rm H,HX}$ depends on how 
the soft excess is modeled).

Many recent forays into modeling soft excesses have found success 
using ionized reflection models 
modified by relativistic smearing, i.e., a blurred
ionized reflector (e.g., Crummy \et\ 2006);
an origin in atomic features could plausibly explain the consistency
of soft excess features across a range of Seyfert black hole masses.
Starting with the best-fit SXPL model, we removed
the soft X-ray power-law and the 6.04 keV emission line,
and added a Ross \& Fabian (2005) reflection
component convolved with the kernel associated with
relativistic motion around a Kerr black hole
(a {\sc Laor} profile). The photon index of the illuminating
continuum was tied to that of the hard X-ray power-law.
The disk outer radius was fixed at 400 $R_{\rm g}$;
the emissivity index was fixed at 3.
The 6.4 keV Fe K$\alpha$ and Ni K$\alpha$ line parameters 
were fixed at the values in the time-averaged spectrum.
The best-fit model had $\chi^2$/$dof$ = 2698.4/1721,                 
with large ($\pm$10$\%$) correlated residuals below $\sim$2 keV
(see Figure 8g).
The best-fit Fe abundance was $<$0.15,
and the best-fit value of the reflector ionization
parameter $\xi$ pegged at the lower limit of 30 erg cm s$^{-1}$,
likely indicating
that lower values of $\xi$ may be more appropriate. 
The low level of Compton reflection $>$ 10 keV in NGC 3227,
$R$ $\lesssim$ 0.5,
as demonstrated in $\S$6, also suggests that
ionized reflection in NGC 3227, if it exists, is not strong.
We do not consider this model further.

Finally, we tried to model the soft excess 
using smeared absorption to model
ionized absorbing material moving at relativistic velocity,
such as a wind launched from the inner accretion disk
{\sc swind1} (Gierli\'{n}ski \& Done 2006).
Assuming a single power-law to model the intrinsic
continuum emission, modified by smeared absorption, 
yielded a poor fit ($\chi^2/dof$ = 8388/1728).
Modeling the intrinsic continuum using two power-laws,
as in the SXPL model, and adding smeared absorption
plus the low-ionization X-ray absorber,
yielded a good fit ($\chi^2/dof$ = 1892/1723), in 
which a smeared ionized absorber with 
log$\xi$ $\sim$2.7 erg cm s$^{-1}$ was able to 
mimic the high-ionization X-ray absorber, 
given the resolution of the pn. 
However, analysis of the RGS spectrum ($\S$5)
supports the need for
both the low- and high-ionization X-ray absorbers, and so
we do not consider this model further.

\section{RGS modeling of the ionized absorbers and emission lines}

We first modeled the continuum of the RGS spectrum
with a model of the same form as the SXPL model from the pn fits:
a steep soft power-law with $\Gamma_{\rm SX}$ $\sim$ 3.2 
absorbed by a cold column $N_{\rm H,local} = 1 \times 10^{21}$ cm$^{-2}$,
and a hard power-law absorbed by a column $N_{\rm H,HX} \sim 1 \times 10^{22}$ cm$^{-2}$,
and dominating the continuum only above $\sim$ 1.5 keV
(and with both $N_{\rm H,HX}$ and $\Gamma_{\rm HX}$ very poorly constrained,
given that the RGS only covers up to 2 keV).
The residuals to this model ($\chi^2$/$dof$ = 1156/433)
are shown in Figure 10. 

There is a broad absorption feature at $\sim$740--780 eV, which we identify as
an Fe M-shell UTA trough. Its energy range (assuming a velocity offset from systemic near zero)
indicates that log$\xi$ of the gas where this feature originates must be in
the range $\sim$ +0.3 to +1.4, according to an {\sc xstar} table; and absorption is mainly due to
species of Fe in the ionization range $\sim$ \ion{}{6}--\ion{}{13}, and likely not due to 
Fe $\sim$ \ion{}{14}--\ion{}{16} (Gu et al.\ 2006)



Many narrow absorption features are evident; the tentative identifications (and lab-frame energies) of
the most visually prominent ones are labeled in Figure 10 and include
\ion{C}{6} Ly$\alpha$ (368 eV), \ion{C}{6} Ly$\beta$ (436 eV),
\ion{N}{7} (500 eV), \ion{O}{8} Ly$\alpha$ (653 eV), \ion{O}{7} He$\beta$ (666 eV; perhaps blended with the \ion{N}{7} edge),
\ion{O}{7} He$\gamma$ (698 eV), \ion{Fe}{17} 3d--2p (812 eV),
\ion{Ne}{9} (922 eV, though this line may
be blended with Fe absorption lines at $\sim$ 918--934 eV), \ion{Ne}{10} (1022 eV) and \ion{Mg}{11} (1352 eV).


A strong edge at 739 eV due to \ion{O}{7} is not obvious, though it could be
blended with the Fe UTA feature. Similarly, an edge at 870 eV
due to \ion{O}{8} is not obvious. The lack of a very strong ($\gtrsim 20-30 \%$ drop) 
\ion{O}{8} edge suggests that any absorbing material containing \ion{O}{8}
has a hydrogen column density $N_{\rm H}$ $<$ 10$^{22.5}$ or so. 
There are also narrow emission features near the energies for 
the \ion{O}{7} and \ion{N}{6} resonance lines (574 eV and 428 eV, lab-frame); these are discussed
further below.

We next modeled the two ionized X-ray absorbers discussed in the pn fits above, using the same
{\sc xstar} table models, with their respective
velocity offsets relative to
systemic initially frozen at zero, and with abundances fixed to solar values. 
$\chi^2$/$dof$ fell to 955.40/429 for values of log$\xi_{\rm lo}$ and log$\xi_{\rm hi}$ near
1.2 and 2.8, respectively, and with column densities for each near 1 $\times$ 10$^{21}$ cm$^{-2}$.
However, many of the absorption lines seemed to be blueshifted slightly.

We tried various combinations of allowing the absolute redshift of each
absorber, $z_{\rm lo}$ and $z_{\rm hi}$, to be thawed from the sytemic value.
Thawing $z_{\rm hi}$ only yielded $\chi^2$/$dof$ = 884.87/428 for a best-fit value of 
$z_{\rm hi}$ near --0.0036 (--0.0075 relative to systemic).
Thawing both $z_{\rm hi}$ and $z_{\rm lo}$, but keeping their values tied, yielded
$\chi^2$/$dof$ = 887.46/428 for a best-fit value of $z_{\rm hi}$ = $z_{\rm lo}$ = 
--0.0033 (--0.0072 relative to systemic).
Finally, thawing both $z_{\rm hi}$ and $z_{\rm lo}$ and allowing them to vary independently,
yielded $\chi^2$/$dof$ = 864.09/427 for best-fit values of 
$z_{\rm hi}$ = --0.0034 (--0.0073 relative to systemic) and
$z_{\rm lo}$ = +0.0016 (--0.0023 relative to systemic).

Finally, to this last model (which had the lowest $\chi^2$ value), 
we added five narrow (width $\sigma$=0.5 eV) Gaussian components to represent emission
at the systemic redshift from the \ion{O}{7} $(f)$, $(i)$ and $(r)$, \ion{N}{7}, 
and \ion{N}{6} $(r)$ (note that we model the sum of the two intercombination lines for \ion{O}{7}).
Their measured energies, intensities, and fit parameters are listed in Table 3.
\ion{N}{6} $(f)$ and \ion{N}{6} $(i)$ emission lines were detected only as upper limits,
with intensities $< 6.7 \times 10^{-5}$ and $< 3.0 \times 10^{-5}$ ph cm$^{-2}$ s$^{-1}$, respectively, and so
are not listed in Table 3.
However, as for the pn fits, we caution that since both warm absorbers require some \ion{O}{7}, \ion{N}{6} and \ion{N}{7} absorption,
the emission line intensities likely have large systematic uncertainties.
Final parameters for the ionized absorbers in our best-fit SXPL model (with $\chi^2$/$dof$ = 785.7/417) are
log$\xi_{\rm lo} = 1.21^{+0.18}_{-0.08}$,
$N_{\rm H,lo} = 1.1^{+0.1}_{-0.2} \times 10^{21}$ cm$^{-2}$, 
log$\xi_{\rm hi} =  2.90^{+0.21}_{-0.26}$, and
$N_{\rm H,hi} = 2.4^{+2.0}_{-1.2} \times 10^{21}$ cm$^{-2}$; other model parameters 
are listed in Table 4. We note that $\Gamma_{\rm SX} = 3.00 \pm 0.25$ and
$N_{\rm H,local} = 10.5^{+1.2}_{-1.9} \times 10^{20}$ cm$^{-2}$, both consistent with the best-fit SXPL model
for the EPIC data.

In this model, the best-fit value for the redshift of the high-ionization absorber is
$z_{\rm hi}$ (absolute) =   $-0.00302^{+0.00057}_{-0.00080}$  or
$z_{\rm hi}$ (relative to systemic) = $-0.00688^{+0.00057}_{-0.00080}$, which corresponds to an outflow velocity
of --(2060$^{+240}_{-170}$) km s$^{-1}$ relative to systemic.
For the low-ionization absorber, 
$z_{\rm lo}$ (absolute) =   $+0.00246^{+0.00144}_{-0.00064}$ or
$z_{\rm lo}$ (relative to systemic) = $-0.00140^{+0.00144}_{-0.00064}$, corresponding to an outflow velocity
of --(420$^{+430}_{-190}$) km s$^{-1}$ relative to systemic.
Given the best-fit values of the two warm absorbers here and the relative 
line strengths expected in those cases, the outflow velocity for the 
high-ionization absorber is constrained primarily by the narrow absorption 
lines due to \ion{Ne}{9}, \ion{Ne}{10}, and \ion{Mg}{11}; constraining the 
respective energies of each of these lines separately yielded corresponding 
outflow velocities relative to systemic of $1300^{+650}_{-300}$, 
$600\pm600$, and $1800\pm600$ km s$^{-1}$, respectively. Isolating 
the observed \ion{C}{6} Ly$\beta$, \ion{N}{7} and \ion{O}{8} Ly$\alpha$
absorption lines yielded similar velocity offsets ($2700\pm700$, 
$1800\pm600$ and $1840\pm460$, respectively), though contributions to 
these lines from both the low- and high-ionization absorber are 
expected, given the best-fit ionization levels.
The only strong absorption features which are unique to only the low-ionization absorber
and unambiguously studied here are the (possibly blended) Fe UTA feature and the 
\ion{O}{7} edge.  Models wherein the outflow
velocity of the low-ionization absorber is either much closer to systemic,
equal to that of the dusty lukewarm UV absorber (tens of km s$^{-1}$; Crenshaw \et\ 2001),
or equal to that of the high-ionization absorber are not definitively ruled out at
high confidence. Setting the absolute value of $z_{\rm lo}$ equal to systemic,  
+0.00353 (--0.00033 relative to systemic, corresponding to an outflow
velocity of 100 km s$^{-1}$), or $z_{\rm hi}$
or +0.003859 and refitting yielded increases in $\chi^2$ of only 9, 6, and 16 respectively.
In each case, the outflow velocity for the high-ionization absorber
remained virtually unchanged from the best-fitting SXPL model.

As an additional confirmation on the {\sc xstar} table results,
we measured the $EW$s of selected individual absorption features.
The measured $EW$ for the Fe UTA (--(13$^{+4}_{-8}$)) eV is consistent with the prediction from
an {\sc xstar} table with $N_{\rm H,lo} = 1.1 \times 10^{21}$ cm$^{-2}$ and log$\xi_{\rm lo}$=1.21.
The measured $EW$s for \ion{Ne}{9} ($-5.5\pm2.0$ eV), \ion{Ne}{10} ($-4.4\pm1.6$ eV) and \ion{Mg}{11} ($-4.1\pm2.0$ eV) are
in good or reasonable agreement (given the data quality)
with the predicted $EW$s from an {\sc xstar} table with 
$N_{\rm H,hi} = 2.4 \times 10^{21}$ cm$^{-2}$ and log$\xi_{\rm hi}$=2.90 (these lines are not expected 
to be generated with significantly noticeable $EW$ in the low-ionization absorber).
The measured $EW$s for \ion{C}{6} Ly $\beta$ ($-1.0\pm0.9$ eV), \ion{N}{7} (--(1.1$^{+1.3}_{-0.2}$) eV) and \ion{O}{8} ($-3.3\pm0.7$ eV)
are in good or reasonable agreement (given the data quality)
with the sums of the predicted $EW$s from the two {\sc xstar} tables.

Most of the remaining data/model residuals are consistent with RGS calibration uncertainties or
are instrumental in nature (Pollack \et\ 2007; observed energies listed): a small dip near 394 eV;
large residuals near the O instrumental edge, 524--530 eV;
a small dip at 573--577 eV (which further adds to the systematic uncertainties associated
with \ion{O}{7} emission lines intensity); 
residuals at $\sim$680--695 eV, associated with an instrumental fluorine feature;
an apparent edge at 850 eV; and
a dip at 937--943 eV. 
Many of these instrumental features are also seen in other RGS spectra of Seyferts (e.g., Braito \et\ 2007).

We rely on the RGS primarily for determination of
warm absorber and soft X-ray emission line parameters;
all hard X-ray parameters are best constrained by the EPIC data.
Still, for the SXPL model, most best-fit parameters (including those for the warm
absorbers) were consistent at the $\Delta\chi^2 = 2.71$ significance level between the RGS and the pn;
$N_{\rm H,HX}$ and $\Gamma_{\rm HX}$ were the exceptions,
and clearly better constrained by the EPIC. 
Simultaneous RGS/pn fitting using the SXPL model ($\chi^2$/$dof$ = 3220/2161) 
yielded best-fit values for $\Gamma_{\rm SX}$, $N_{\rm H,local}$ 
and warm absorber column densities and ionization parameters
consistent with those measured by the RGS and EPIC-pn separately.
Similarly, modeling the continuum in the RGS data
using a blackbody or a {\sc CompST} component yielded good fits
($\chi^2$/$dof$ = 805.7/417 and 806.4/416, respectively), with
a best-fit blackbody temperature and {\sc CompST} parameters consistent with
those measured using only EPIC data, and
with no significant changes to the warm absorber or emission line parameters
compared to using a soft power-law component to model the soft excess.

To this point, we have been assuming that the ionized absorbers along the
line of sight have been full-covering. Using the RGS data, we 
tested a model wherein only a fraction of the continuum radiation
is absorbed by the ionized absorbers. We found the upper limit to
the fraction of continuum radiation that remains unattenuated by
ionized absorption to be 30$\%$, i.e., a covering fraction of $>70\%$      
for the RGS data. For the pn data, the covering fraction is inferred
to be $>93\%$.

We also tested for the presence of radiative recombination continuum
(RRC) features in the RGS data associated with
H- or He-like ions, expected if the emission lines are associated with gas that is
photo-ionized. Using a {\sc redge} component in {\sc xspec} with
energy fixed and with $k_{\rm B}T$ fixed at 
0.5 keV (arbitrary value), we found upper limits in the range 5--8 eV for 
\ion{C}{5}, \ion{C}{6}, \ion{N}{6} and \ion{O}{7} RRCs, upper limits  $\leq$ 20 eV for \ion{O}{8} and 
\ion{Ne}{9}, and an upper limit near 50 eV for an \ion{Ne}{10}
RRC. A \ion{N}{7} RRC at 667 eV significantly improved the fit, but
that was due to fitting the instrumental fluorine feature,
and therefore is likely not real.
Finally, we note no significant emission from the \ion{Fe}{17} L line (3d--2p) 
at 826 eV (upper limit of 1 eV), limiting the possibility that the observed
emission lines are generated via collisional ionization processes.

We now discuss the search for X-ray spectral features in the RGS
associated with dust embedded in the warm absorbers.
The expected effects of embedded dust on the X-ray spectrum
are: 1) Edges due to neutral Fe, O, C, Si and Mg
(depending on dust composition) in excess from the absorption expected from gas.
The O edge would be at 531 eV. The three Fe edges are the L3, L2, and L1 edges,
at 707, 721 and 846 eV (assuming oxidized Fe; edges due to pure Fe would be $\sim$3 eV higher), 
roughly in a 10:5:1 ratio (e.g., Kortright \& Kim 2000; Bearden \& Burr 1967; Schulz \et\ 2002).
2) There may be weak, narrow absorption lines due to Fe oxide species, in
particular at 702 eV (e.g., Crocombette \et\ 1995). 
3) Fe, C and O abundances in the warm absorber would be
low due to depletion onto dust grains (Snow \& Witt 1996; Komossa \& Fink 1997a).


Near 707 eV and 721 eV (the Fe L3 and Fe L2 edge energies)
the residuals in the RGS spectrum of NGC 3227
do not appear obviously edge-like.
Inserting an edge at any of neutral Fe or O energies does not improve the
fit; upper limits to the optical depths of O, Fe L3 and Fe L2 in excess of the neutral absorption
associated with the gas components modeled above (both local and Galactic) 
are 0.17, 0.05 and 0.21, respectively.
In contrast, the Fe L3 edges seen in Cyg X-1 (Schulz \et\ 2002) and MCG--6-30-15 (Lee \et\ 2001)
with the {\it Chandra} HETGS are both $\gtrsim$50$\%$ drop across the edge ($\tau \gtrsim 0.7$).
From the Fe L3 edge in NGC 3227, the implied upper limit on the \ion{Fe}{1} column density is
$\sim10^{17}$ cm$^{-2}$.
Assuming the abundances of Lodders (2003), the corresponding
equivalent hydrogen column density is constrained to be less than
$\sim10^{21.5}$ cm$^{-2}$. This limit is consistent with the 
equivalent hydrogen column density implied by optical line absorption (see $\S$1).
The limit on the $EW$ (relative to locally-determined and unabsorbed continuum) of
a narrow (width $\sigma$ = 0.5 eV) absorption line at 702 eV is --0.5 eV. 
Finally, thawing the abundances of Fe, O, and Mg in the warm absorbers
(abundances in both absorbers tied) does not yield significant evidence for
deviations from solar abundances.


\section{Constraining the $>$10 keV continuum with RXTE and Swift-BAT}

We fit archival {\it RXTE} data and {\it Swift} Burst Alert Telescope (BAT) 9-month survey
data to constrain 
the amount of the Compton reflection and constrain any high-energy power-law cutoff.
We did not do simultaneous fitting with the EPIC spectrum,
since the {\it XMM-Newton} and latest {\it RXTE} and BAT data obtained
were separated by a year and since
the only two broadband model components in the range of overlap
are the variable hard X-ray power-law and the hard X-ray absorber,
whose column $N_{\rm H,HX}$ $\lesssim 10^{21}$ cm$^{-2}$ is not well constrained by the PCA.
However,  the observed 2--10 keV flux during the 2006 {\it XMM-Newton} observation, 
$F_{2-10}$ = 3.5$\times$10$^{-11}$ erg cm$^{-2}$ s$^{-1}$, fell in the approximate range 
probed by the {\it RXTE} monitoring; only during MJD $\sim$51850--52050
was $F_{2-10}$ observed to be consistently below 2$\times$10$^{-11}$ erg cm$^{-2}$ s$^{-1}$. 
This suggests that {\it XMM-Newton} caught the source in a ``typical'' flux state.

We fit the PCA data over the 3.3--30 keV range; the source is
not well detected at higher energies.
HEXTE data below 24 keV were ignored as the responses of 
clusters A and B diverge somewhat for faint sources (N.\ Shaposhnikov, 
private communication, 2005); there is good agreement above this energy. 
HEXTE data above 100 keV were also ignored. 
14--195 keV BAT data were included; four-channel spectra and response matrices
were taken from the BAT 9-month AGN catalog 
website\footnote{http://swift.gsfc.nasa.gov/docs/swift/results/bs9mon/}.
A constant coefficient was included in all spectral models
to account for minor normalization offsets between the PCA and HEXTE; its value for
the HEXTE spectra relative to the PCA spectrum was typically 0.8--0.9.
No constant was needed for the BAT spectrum relative to the PCA spectrum.

We applied a model consisting of a power-law (initially with a high-energy cutoff
set arbitrarily at $E_{\rm c}$ 500 keV),
a Gaussian emission line at 6 keV to model Fe K$\alpha$ emission,
and a component to model Compton reflection from neutral material, using
{\sc pexrav} (Magdziarz \& Zdziarski 1995), all modified by
a cold column $N_{\rm H,HX}$ constrained to be less than $6 \times 10^{21}$ cm$^{-2}$.
The {\sc pexrav} component's inclination was fixed at 45$\degr$, solar Fe abundances were assumed,
and the high energy cutoff was tied to that of the power-law.
The best-fit model had $\chi^2$/$dof$ = 297/111, the
photon index $\Gamma_{HX}$ was 1.66$\pm$0.01, and the 
value of the Compton reflection strength 
$R$\footnote{$R \equiv \Omega/2\pi$, where $\Omega$ is the solid angle subtended by the reflector} 
was $0.16^{+0.08}_{-0.02}$. Allowing the high-energy cutoff to be free and re-fitting,
$\chi^2$/$dof$ fell to 222/110 (significant at 5.6$\sigma$ confidence according to an $F$-test).
$\Gamma_{HX}$ was 1.63$\pm$0.02, $R$ was 0.40$\pm$0.07, and $E_{\rm c}$ = 90$\pm$20 keV.
The cold column $N_{H,HX}$ was $6.0^{+0}_{-0.8} \times 10^{21}$ cm$^{-2}$
(uncertainty pegged at upper limit).
The residuals to the best-fit models with and without a high-energy cutoff
are shown in Figure 11\footnote{The presence of the high-energy cutoff is also inferred from fitting PCA+HEXTE
only: excluding the BAT data, and refitting, we find that including a high-energy cutoff 
still significantly improves the fit, with best-fit values consistent with the
PCA + HEXTE + BAT fit, though the statistical uncertainty on $E_{\rm c}$ is 1.8 times larger.
Ignoring the HEXTE data, and fitting PCA + BAT only,
the best-fit values of $\Gamma_{HX}$, $R$, and $E_{\rm c}$ are 1.62$^{+0.02}_{-0.04}$, 
0.53$^{+0.19}_{-0.08}$ and $65^{+10}_{-25}$ keV, respectively.}.

Because the {\it RXTE} and {\it XMM-Newton}
observations were not simultaneous, it is not clear whether the
slightly flatter value of $\Gamma_{\rm HX}$ for the pn in the SXPL model (1.57$\pm$0.02) 
compared to that for the time-averaged {\it RXTE} spectrum 
is due to source variability or associated with systematic differences in
modeling spectra taken with two instruments with very different energy resolution. 
For example,
the average Fe K$\alpha$ line flux as measured by the PCA was $9 \pm 1 \times 10^{-5}$ ph cm$^{-2}$ s$^{-1}$,
but this is not automatically 
an indication of line variability between the {\it RXTE} and {\it XMM-Newton}
observations given the differences in PCA and EPIC resolution, the
systematic uncertainties associated with fitting such a complex model to the low-resolution
PCA data, and possible uncertainties associated with PCA/pn cross-calibration.
All uncertainties listed here are
one-dimensional statistical errors only and do not account
for systematic uncertainties associated with correlations between
parameters or for systematic instrumental uncertainties associated
with observing faint objects.
However, we can safely conclude that the value of $R$ must be relatively low:
Figure 12 shows contour plots for $E_{\rm c}$ and $R$ as a 
function of $\Gamma_{HX}$ from the joint {\it RXTE}+{\it Swift}-BAT fit, 
demonstrating that values of $R$ $>$ 0.6 are rejected at high confidence.

\section{Time- and Flux-Resolved Spectral Fits}

We investigated the temporal behavior of the X-ray spectrum of NGC 3227 on
a range of time scales.  Any rapid variation in ionization parameter
would rule out models where the warm absorber is spatially extended
and/or located very far from the black hole;
tracking ionization responses to continuum flux variations
can yield constraints on the location of the warm absorbers
(e.g., Krongold \et\ 2007).   
Similarly, any rapid variation in the strength of an Fe K emission feature
(e.g., Tombesi \et\ 2007) would suggest an origin close to the X-ray illuminating source.
We used the {\it XMM-Newton} spectrum to study variability on time scales $\lesssim$ 1 day
(hereafter ``short'' time scales);
we used the {\it RXTE} intensive monitoring from 2000 April to June to investigate
variability on time scales from $\sim$2 weeks to 2 months (hereafter ``medium'' time scales),
and the 1999--2005 monitoring to investigate variability on time scales from
months to several years (``long'' time scales'').

\subsection{{\it XMM-Newton} time-resolved spectral fits}

We investigated the $<$1-day time-resolved spectral behavior 
of NGC 3227 during the {\it XMM-Newton} observation
by first splitting the EPIC-pn data into five segments
of equal duration (good exposure time 18 ks); shorter durations would not have yielded
adequate photon statistics in narrow features.
In the Fe K bandpass, spectral fitting revealed several
``candidate'' narrow transient emission features between 4.5 and 5.2 keV,
each with $\Delta$$\chi^2$ in the range --11 to --14.
Monte Carlo simulations (see $\S$3) showed that each of these emission
features were significant at 92--97$\%$ confidence.
However, when one takes into consideration the number of segments searched over
(Vaughan \& Uttley 2008),
the significance levels drop to 60--85$\%$ confidence; these features are
not discussed further. Considering the broadband spectrum, all 
parameters for the neutral and ionized absorbers and the \ion{O}{7}
emission line were consistent with those found in the
time-averaged spectrum. 
Similarly, flux-resolved spectral analysis,
performed on high- and low-flux pn spectra (extracted above and 
below a 0.2--12 keV pn count rate of 11.0 ct s$^{-1}$, respectively), 
yielded no significant evidence for variability of the neutral or 
ionized absorbers, the 6.4 keV Fe K$\alpha$ line, or the \ion{O}{7} line.
The 6.04 keV line was not significantly detected with these
short exposures.

However, we noticed in these time-resolved fits that
the normalization of the soft excess increased by $\sim$25$\%$
from the first 20 ks segment to the second segment.
Similar results were obtained when modeling the soft excess as
a steep power-law, a blackbody, or a {\sc CompST} component.

The rapid (ks and shorter) variations seen simultaneously in both the
hard and soft X-ray light curves (Figure 1) are similar to those routinely
detected in other Seyferts; the associated variability 
mechanisms will be explored in depth in a future paper (Ar\'{e}valo 
et al., in prep) and are not discussed further here. 
Here, we are concentrating on the soft X-ray band's 
relatively slower and distinct trend. 
Since this trend occurs over tens of ks, we will refer to it
as "rapid" for the remainder of this paper.

We repeated the time-resolved analysis, dividing the data into
ten segments of equal duration (9 ks good exposure time), fitting
the SXPL model, and
keeping all warm and cold absorber, \ion{O}{7} line
and Fe emission line parameters frozen at their time-averaged values.

The resulting light curves of $\Gamma_{\rm HX}$, $\Gamma_{\rm SX}$ 
and the normalizations of the SXPL and HXPL components 
are plotted in Figure 13. The SXPL normalization 
seems to track the light curve of 0.2--1 keV flux in Figure 1.
To further illustrate the rapid variability of the soft excess, the
data for segments 1 (the lowest soft excess flux bin), 2, and 6 
(the highest soft excess flux bin) are plotted in Figure 9.

Finally, fractional variability amplitudes ($F_{\rm var}$
see Vaughan \et\ 2003 for a definition) were calculated for 
16 energy bands. The resulting $F_{\rm var}$ for the entire duration, for
the first 20 ks, and for the final 80 ks are plotted in Figure 14,
further illustrating the rapid variability in the soft excess.
Further detailed variability analysis of the variable X-ray continuum,
including coherence, intra-X-ray time lags and power spectral
density function measurement, will be presented in a future 
paper (Ar\'{e}valo et al., in prep).

\subsection{{\it RXTE} time-resolved spectral fits}

Time-resolved spectral fitting of the {\it RXTE} data
closely followed Markowitz, Edelson \& Vaughan (2003).
Bin sizes were chosen to optimize the
trade off between minimizing the uncertainties on
Fe line flux and maximizing the number of bins.
This yielded 5 bins, each of duration 13 days, for the medium
time scale. On the long time scale, start/stop times were 
chosen to avoid 60-days gaps where {\it RXTE} did not observe the
source due to sun-angle constraints, with most
durations roughly 100 days before loss of PCU0 on 2000 May 12, 
and 150 days afterward, yielded 16 bins. 
Response matrices were generated separately for each segment.

We used a model consisting of a power law component, 
a Gaussian component to model Fe K emission, and
a {\sc pexrav} component to model Compton reflection,
all absorbed by cold material (with a column density $N_{\rm H, HX}$
constrained to be $ < 6 \times 10^{21}$ cm$^{-2}$,
unless the bins included data taken during the 2000-1 obscuring event).
In the best fit models, the best-fit values of $R$, Fe line rest-frame
energy centroid, and Fe line width $\sigma$ were
$0.4^{+0.2}_{-0.1}$ ($0.2 \pm 0.1$),
$6.26^{+0.10}_{-0.11}$  ($6.27^{+0.05}_{-0.04}$) keV, and
$0.50^{+0.32}_{-0.14}$  ($0.33^{+0.09}_{-0.08}$) keV
in the medium (long) time scale fits, respectively.
To determine which parameters to leave free during each
time-resolved fit, we first fit all segments simultaneously with
all parameters tied, and thawed one parameter at a time,
testing for significant improvement in the fit according to an $F$-test.
We found it was significant to thaw the power law normalization, $\Gamma_{\rm HX}$,
Fe line intensity $I_{\rm FeK\alpha}$, and $N_{\rm H,HX}$ in the individual,
time-resolved fits.




Errors for $I_{\rm FeK\alpha}$ and $\Gamma_{\rm HX}$
were derived using the point-to-point 
variance\footnote{See $\S$3.3 of Markowitz, Edelson \& Vaughan 
(2003) for further details and the definition of the point-to-point variance}.
Errors on values of $F_{2-10}$ within a time bin 
were determined from the 2--10~keV continuum light curve,
using the mean flux error on the $\sim$1~ks exposures in that time bin.
Figure 15 shows the resulting light curves for $F_{2-10}$,
$I_{\rm FeK\alpha}$, and $\Gamma_{\rm HX}$.

Fractional variability amplitudes $F_{\rm var}$ were calculated.
For the medium time scale, $F_{\rm var}$ for
the $F_{2-10}$ and $I_{\rm FeK\alpha}$ light curves were
38.4$\pm$0.3$\%$ and 22.7$\pm$10.3$\%$, respectively,
although we caution that with only 5 data points, these measured
values of $F_{\rm var}$ are likely not highly reliable.
For the long time scale, 
$F_{\rm var}$ for $F_{2-10}$ and $I_{\rm FeK\alpha}$ were
32.7$\pm$0.7$\%$ and 20.0$\pm$4.3$\%$.
Qualitatively, these 
results are consistent with what Markowitz, Edelson
\& Vaughan (2003) found for a small sample of Seyferts:
the Fe line does not vary as strongly as the continuum flux.


Inspection of the $F_{2-10}$ and $I_{\rm FeK\alpha}$
light curves would seem to indicate, to the human eye at least,
similar variability trends. However, with so few points 
(especially for the medium time scale), 
such a statement is not highly significant. 
On the medium time scale, the best-fit linear relation is 
$I_{\rm FeK\alpha} = 2.05 \pm 1.31 \times F_{2-10} +  4.1 \pm 3.1$,
with $I_{\rm FeK\alpha}$ in units of 10$^{-5}$ ph cm$^{-2}$ s$^{-1}$ and
$F_{2-10}$ in units of $10^{-11}$ erg cm$^{-2}$ s$^{-1}$.
Figure 16 shows the zero-lag correlation plot of
$I_{\rm FeK\alpha}$ as a function of $F_{2-10}$,
with the best-fit linear relations plotted as dashed lines.
The zero-lag Spearman rank correlation coefficient\footnote{calculated at http://www.wessa.net/rankcorr.wasp} is 0.700; the corresponding null hypothesis
probability (the probability that the
correlation could arise from randomly chosen data points)
is 16$\%$; i.e., the correlation is significant only at 84$\%$
confidence. 

For the long time scale data, the best-fit linear relation is 
$I_{\rm FeK\alpha} = 1.44 \pm 0.30 \times  F_{2-10}  + 3.5 \pm 1.2 $.
The zero-lag Spearman rank correlation coefficient is 
r=0.739, significant at 99.6$\%$ confidence. 
We searched for lags using an Interpolated Cross Correlation Function
(ICF; Gaskell \& Peterson 1987, White \& Peterson 1994), with errors determined
using the bootstrap method of Peterson et al.\ (1998).
The ICF peak correlation coefficient $r_{max}$=0.854
was reached at a delay of 75 $\pm$ 690 days ($F_{2-10}$
leading $I_{\rm FeK\alpha}$), i.e., consistent with 
zero lag. 

With such few data points and low correlation  
coefficients on both medium and long time scales, 
any claim of a correlation must be deemed tentative at best.
Cross correlation analysis is further complicated by the red-noise nature 
of the continuum (and likely line) light curves. Specifically, 
cross-correlation between two unrelated red-noise
light curves can randomly yield spurious correlations
with higher than expected values of the
correlation strength $r_{\rm max}$ (e.g., Welsh 1999).
Lags where $r_{\rm max}$ is not very close to 1.0
should thus be treated with skepticism.
Additional monitoring, spanning much longer durations and
wider ranges in both $F_{2-10}$ and $I_{\rm FeK\alpha}$ and encompassing 
additional upward/downward trends in the light curves
of both $F_{2-10}$ and $I_{\rm FeK\alpha}$ are required to 
critically test for any significant correlation.
We therefore conclude that there is, at best, tentative
evidence from the {\it RXTE} time-resolved spectral fits for a significant
fraction of the the Fe line intensity to respond to continuum 
variations on time scales shorter than 700 days.
The lack of strong hard X-ray variability during the
{\it XMM-Newton} observation ($F_{\rm var} = 8.5\pm0.2\%$ for the
3--10 keV light curve, binned to 600 s)
means we cannot draw any conclusions about the response
of the line on $\lesssim1$ day time scales.


\section{X-ray/UV continuum light curve correlations}


The OM light curve of NGC 3227 displayed in Figure 1 shows a 
$\sim$10$\%$ increase across the observation, with an RMS of 
2.8$\%$. Rapid optical/UV variability in Seyferts on time scales
of $\lesssim$ 1 day is relatively rare. Of the sample of 8 
Seyferts examined by Smith \& Vaughan (2007), NGC 3783 displayed 
the strongest variability on $\lesssim$ 1 day time scales, with 
an RMS of 2.9$\%$ in the UVW2 filter; there were only 
3 additional observations using either the UVW2 or U filter
where the RMS was $> 1.8\%$.
The current {\it XMM-Newton} data would thus seem to indicate that
NGC 3227 is towards the top of the list of Seyferts which display
strong UV continuum variability on $\lesssim$ 1 day time scales.
In NGC 3227, the X-rays are much more strongly variable on these
time scales (RMS for the 0.2--1 and 3--10 keV light curves
were 15.8$\%$ and 8.9$\%$, respectively), a result similar to 
what Smith \& Vaughan (2007) found for their sample.

Visually, it is tempting to connect the gradual brightening 
observed in the OM light curve with that observed simultaneously
in the soft X-ray band during the {\it XMM-Newton} observation. 
However, the rise in soft X-ray flux ($\sim$40$\%$) is much greater
than that observed in the UV band ($\sim$10$\%$).
We calculated ICFs and Discrete Correlation Functions (DCF; Edelson \& Krolik 1988) between the UV and soft X-ray light curves (the latter rebinned to 1400 s); the results are 
plotted in Figure 17, along with 90 and 95$\%$ confidence limits
from the Bartlett method. We found no significant correlation:
$r_{\rm max}$ peaks at only $\sim$0.5; there are no ``bends'' or
multiple trends in the OM light curve to drive a correlation.
The warning regarding cross-correlation analysis performed on two
unrelated red-noise light curves bear repeating:
beware of spuriously high values of $r_{\rm max}$ (e.g., Welsh 1999)
and treat values of $r_{\rm max}$ which are not very close to 1.0
with skepticism. The suggestion of a correlation between the 
soft X-ray band and the UV band, while visually tempting, is 
thus speculative at best.

\section{Discussion}

We have analyzed a $\sim$100 ks {\it XMM-Newton}
long-look of NGC 3227, observed in December 2006. 
In both the EPIC-pn and RGS spectra, we have modeled the ionized 
X-ray absorber using two components, with very similar column
densities, and  
with ionization parameters log$\xi$ near 1.2 and 2.9 (in the RGS spectrum),
With the RGS, we have constrained
the outflow velocity of the high-ionization absorber to be
--(2060$^{+240}_{-170}$) km s$^{-1}$ relative to systemic.
The best estimate for the outflow velocity relative to systemic for the low-ionization
absorber is --(420$^{+430}_{-190}$) km s$^{-1}$,
though a wider range of possible outflow velocities
cannot be ruled out at very high confidence.
In $\S$9.1, we discuss further details of these ionized outflows,
including exploring connections to the low-velocity, very low ionization,
dusty lukewarm absorber observed in the UV.

The steep soft excess, which does not seem be to affected by the same neutral material
obscuring the hard X-ray continuum, is shown with the EPIC pn to be rapidly variable
compared to other Seyferts on $\lesssim$ 1 day time scales; its
normalization increases by $\sim$25$\%$ in $\sim$20 ks.
The UV continuum light curve during the {\it XMM-Newton}
observation also shows a relatively strong 
increase compared to most other Seyferts on $\lesssim$ 1 day time scales.
Possible origins for these behaviors are discussed in $\S$9.2.

The hard X-ray power-law continuum in both the {\it XMM-Newton} pn and MOS
fits and in the PCA + HEXTE + BAT fits is somewhat low, $\Gamma_{\rm HX} \sim 1.5-1.6$.
Using accumulated {\it RXTE} PCA and HEXTE archival data from
over 6 years of monitoring, plus {\it Swift}-BAT 9-month survey data,
we constrain the strength of the Compton reflection component $R$
to be $\lesssim$ 0.5. We also find the first evidence 
for a high-energy continuum cutoff in this source, at 90$\pm$20 keV.
The high-energy continuum is discussed further in $\S$9.3.

The narrow Fe K$\alpha$ emission line at 6.4 keV is resolved in the pn
spectrum. From time-resolved spectral fits to the {\it RXTE}-PCA
data, we find tentative evidence for a significant fraction
of the Fe K line flux to track variations in
the continuum flux $F_{2-10}$ on time scales of $<$700 days. The 6.4 keV
Fe K$\alpha$ line properties are discussed further in $\S$9.4.

In addition, we find significant evidence for emission near 6.0 keV 
in both the pn and MOS 1+2 spectra.
It is modeled approximately equally well as a narrow 
emission line or as the red wing to a relativistically-broadened Fe K emission line,
as discussed in $\S$9.5.

\subsection{Overview of the X-ray absorbers}

In the {\it XMM-Newton} RGS spectrum, we model two zones of
absorption, with log$\xi_{\rm lo} = 1.21^{+0.18}_{-0.08}$ and log$\xi_{\rm hi} = 2.90^{+0.21}_{-0.26}$, and with very
similar column densities, 
$N_{\rm H,lo} = 1.1^{+0.1}_{-0.2} \times 10^{21}$ cm$^{-2}$ and
$N_{\rm H,hi} = 2.4^{+2.0}_{-1.2} \times 10^{21}$ cm$^{-2}$.
Other AGN have been reported to host multiple
warm absorber components with different
ionization parameters but similar hydrogen column densities
(see e.g., Blustin \et\ 2005).
The blueshift relative to systemic of the high-ionization absorber was significantly constrained, yielding
an outflow velocity relative to systemic of --(2060$^{+240}_{-170}$) km s$^{-1}$.
The best-fit blueshift relative to systemic for the low-ionization absorber corresponded to
an outflow velocity relative to systemic of --(420$^{+430}_{-190}$) km s$^{-1}$;
however, as this is primarily constrained by only two (possibly blended) absorption features,
models with blueshifts ranging from zero to identical to that for
the high-ionization absorber were not clearly rejected.
It is therefore not clear if the two zones are physically and kinematically distinct or
if there exists a single ionized absorber spanning a broad 
range of ionization levels, a possibility to consider if
the ionized X-ray absorbing gas is spatially extended.
For instance, Gon\c{c}alves \et\ (2006) model the warm absorbing layers in NGC 3783
as a single constant-density gas component in pressure equilibrium.

To estimate the distance $r$ between the central black hole
and the outflowing gas in NGC 3227, we can use $\xi = L_{\rm ion}/(nr^2)$,
where $n$ is the number density, and $L_{\rm ion}$ is the 1--1000 Ryd 
illuminating continuum luminosity. We estimate the maximum possible
distance to the material by assuming that the thickness
$\Delta$$r$ must be less than the distance $r_{\rm max}$.
The column density $N_{\rm H}$ = $n$$\Delta$$r$,
yielding the upper limit $r_{\rm max}$ $<$ $L_{\rm ion}$/($N_{\rm H}$$\xi$).
We estimate the unabsorbed 1--1000 Ryd flux to be
$\sim 3 \times 10^{-10}$ erg cm$^{-2}$ s$^{-1}$, which  
corresponds to $L_{\rm ion} \sim 1.5 \times 10^{43}$ erg s$^{-1}$
(using a distance of 20.3 Mpc, and assuming
$H_{\rm o}$ = 70 km s$^{-1}$ Mpc$^{-1}$ and $\Lambda_{\rm o}$ = 0.73).
Assuming two physically distinct warm absorbers, 
for the high-ionization X-ray absorber, this yields
$r_{\rm max}$ = 3.6 pc. For the low-ionization X-ray absorber, 
the constraints are even weaker: $r_{\rm max}$ = 150 pc. 
These constraints are much poorer compared to that derived by Gondoin \et\
(2003) for a one-zone absorber, $r_{\rm max} = 0.45$ pc, as those authors found a higher value
for the column density and a lower value for the ionizing flux.

However, we can also derive a minimum radial distance from the black hole $r_{\rm min}$ for the winds
via the requirement for the outflow velocity $v$ to be 
greater than the escape velocity: For the high-ionization X-ray absorber, $r_{\rm min} = (c^2/v^2)R_{\rm g} = 
17000 R_{\rm g} = 40$ light-days, placing it outside
both the BLR and the inner radius of the dust as
reverberation-mapped by Suganuma \et\ (2006).


Under the assumption that the gas is in equilibrium and that
the outflow velocity is a constant, the mass outflow rate 
\.{M}$_{\rm out}$ of the X-ray absorbers can be derived via 
conservation of mass:
\.{M}$_{\rm out}$ = $\Omega$$n$$r^2$$v$$m_{\rm p}$,
where $v$ is the outflow velocity, $m_{\rm p}$ is the proton mass,
and $\Omega$ is the covering fraction.
We then substitute $n$$r^2$ = $L_{\rm ion}$/$\xi$.
Assuming $\Omega$ = 0.3 (arbitrary), we find, for the high-ionization absorber,
\.{M}$_{\rm out}$ $\sim 2 \times 10^{24}$ gm s$^{-1}$ $\sim$ 0.03 $\Msun$ yr$^{-1}$. 
For the low-ionization absorber, assuming $v = 420$ km s$^{-1}$,
\.{M}$_{\rm out}$ $\sim 2 \times 10^{25}$ gm s$^{-1}$ $\sim$ 0.3 $\Msun$ yr$^{-1}$.
We note that the actual outflow rate should be lower if there is an 
extreme degree of collimation along the line of sight.
We can compare the outflow rate
to the inflow accretion rate \.{M}$_{\rm acc}$ using
$L_{\rm bol}$ = $\eta$\.{M}$_{\rm acc}$$c^2$,
where $\eta$ is the accretion efficiency parameter, typically 0.1.
Woo \& Urry (2002) estimate $L_{\rm bol}$ for NGC 3227
to be $7.2 \times 10^{43}$ erg s$^{-1}$, which means an accretion rate
relative to Eddington, $L_{\rm bol}/L_{\rm Edd}$, of 1.4$\%$.
For $\eta=0.1$, \.{M}$_{\rm acc} =  0.01 \Msun$ yr$^{-1}$.
The kinetic power associated with the outflow component,
estimated as \.{M}$_{\rm out}$$v^2$, is thus in the approximate range
$10^{40-41}$ erg s$^{-1}$.

The outflow mass rate and kinetic energy are rough estimates only,
but it does appear likely that the outflows represent at least a large fraction
of the AGN's accretion rate. If the X-ray ionized outflows
are long-lived, then a sustained feeding of the black hole would be difficult.


\subsubsection{Helium-like emission line diagnostics}

It is not obvious whether the emission lines detected in the
RGS spectrum are due to collisional ionization or
photo-ionization. We do not detect a strong \ion{Fe}{17} L 
3d--2p $^1$P$_1$ line at 826 eV, an indicator of collisional ionization.
However, RRC lines, indicators of photo-excitation, are not obvious, either.

We can attempt to
use diagnostics associated with the helium-like O and N emission triplets.
In a collisionally- (photo-) ionized plasma, the resonance (forbidden) 
line dominates. However, in NGC 3227, there is not
significant dominance of one of these lines in either emission triplet.
The presence of helium-like absorption lines may be a factor: 
the emission and absorption lines are  
not completely separated in energy given the modeled blueshift of
the absorption lines, the RGS resolution and the signal/noise ratio
of the RGS data. This fact also complicates our use of the density
indicator $R \equiv f/i$ and the temperature indicator $G \equiv (f+i)/r$
(e.g., Porquet \& Dubau 2000)
where $f$, $i$ and $r$ are the intensities of the
forbidden, intercombination, and resonance lines, respectively
($i$ is the summed intensity of both intercombination lines).
As $G$ depends on $r$, it can indicate whether photo-ionization
dominates or whether a photo-/collisional-ionization hybrid is
applicable (Porquet \& Dubau 2000).
Porter \& Ferland (2007) warn that although $G$ indicates temperature in
a collisionally-ionized plasma, it should not be used as a temperature
indicator in a photo-ionized plasma (in addition, see Porter \& Ferland 2007 for
warnings regarding usage of $R$ as density indicator  in photo-ionized plasmas).

For the \ion{O}{7} emission triplet in NGC 3227, we measure 
$G = 4.7^{+9.3}_{-4.0}$; $G < 2.0$ for the \ion{N}{6} triplet.
Uncertainties here are statistical only and do not include
systematic effects associated with the presence of absorption lines.
These values thus do not yield any useful constraints on whether
collisional- or photo-ionization dominates. 
For the \ion{O}{7} emission triplet, $R = 1.4^{+1.0}_{-0.9}$.
Assuming the plasma is either purely ionized or
a hybrid of collisionally- and photo-ionized material,
and assuming a temperature near $10^6$ K, electron densities 
of roughly $10^{10.5-11.5}$ cm$^{-3}$ are implied
(Porquet \& Dubau 2000). For either collisional or photo-ionization, 
densities above $10^{12}$ cm$^{-3}$ are ruled out by the presence of the
strong \ion{O}{7} $(f)$ line.


\subsubsection{Long-term variability in the cold and ionized absorbers}

Variations in the column density of neutral absorbing gas along the line of sight 
has been reported to occur on a wide range of time scales (days to years)
in both Seyfert 1 AGN (e.g., I Zw 1, Gallo et al.\ 2007; NGC 4151, Puccetti et al.\ 2007;
NGC 3516, Turner \et\ 2008, Markowitz \et\ 2008)
and Seyfert 2 AGN (Risaliti et al.\ 2002, 2005).
NGC 3227 is no exception, given the 2000-1 obscuring event by a lowly-ionized
(log $\xi \sim 0$) dense cloud of column density $2.6 \times 10^{23}$ cm$^{-2}$
and inferred to be located in the BLR (Lamer \et\ 2003).
That cloud is likely physically distinct from the local cold
gas and the ionized X-ray absorbers, as
$N_{\rm H,local}$, $N_{\rm H,WA1}$ and $N_{\rm H,WA2}$
modeled from the 2006 {\it XMM-Newton} observation
are much lower than that for the 2000-1 obscuring cloud.
Nonetheless, there does seem to be evidence for variations in
$N_{\rm H,local}$ on time scales of years in NGC 3227: 
George \et\ (1998b) noted $N_{\rm H,local}$
to increase from $\sim 3 \times 10^{20}$ cm$^{-2}$ in 1993 to $\sim 3 \times 10^{21}$ cm$^{-2}$ by 1995.
Gondoin \et\ (2003)  reported $N_{\rm H,local} = 7 \times 10^{20}$ cm$^{-2}$ from the 2000 {\it XMM-Newton} 
observation; that value is the same as the column here in the 2006 EPIC-pn spectrum.
Assuming that discrepancies in values of $N_{\rm H,local}$ inferred from different missions
are intrinsic to the source and not due to cross-instrumental calibration uncertainties,
the range in measured $N_{\rm H,local}$ values would suggest the
presence of cold gas along the line of sight at $<<$pc radii from the black hole.


George \et\ (1998b) noted an increase in the ionized absorber column density, from 
$\sim 3 \times 10^{21}$ cm$^{-2}$ in 1993 to $\sim 3 \times 10^{22}$ cm$^{-2}$ 1995,
ruling out a location outside the NLR. The value of $N_{\rm H,WA}$ obtained by 
Gondoin \et\ (2003) using {\sc xstar} modeling is $2.7 \pm 0.7 \times 10^{21}$ cm$^{-2}$. 
Variability in $N_{\rm H,WA}$ between 2000 and 2006 is thus implied to be significant only 
at the 2.3$\sigma$ confidence level. Furthermore, Gondoin \et\ (2003) did not include a component
to model a soft X-ray excess; if the soft excess was intrinsically present during the 2000
observation, then their estimate of $N_{\rm H,WA}$ would be too high.
Assuming the variations in $N_{\rm H,WA}$
since 1993 to be real, then at least some fraction of the ionized X-ray absorbing gas is likely present
at $<<$pc radii.

\subsubsection{Possible connections between the X-ray and UV ionized absorbers}

The extended dusty lukewarm absorber (hereafter DLWA) studied by Crenshaw \et\ (2001) 
may be a physically and/or kinematically distinct component from the ionized X-ray absorbers. 
Crenshaw \et\ (2001) report the ionization parameter $U$ of the DLWA
to be 0.13 (see George \et\ 1998a for definitions of the optical/UV ionization parameter
$U$ and the X-ray ionization parameter $U_{\rm x}$). 
Assuming an optical-to-X-ray spectral index $\alpha_{\rm ox}$ (estimated from
photometric measurements taken from NED), and using the conversions from
Figure 1 of George \et\ (1998a), $U/U_{\rm x} \sim 250$ and 
$\xi/U_{\rm x} \sim 5000$, we can translate our best-fit ionization parameters
for the low- and high-ionization X-ray absorbers,
log$\xi$=1.21 and log$\xi$=2.90, respectively, 
into corresponding values of $U = 1.0$ and 40. 
The values of the ionization parameters for the DLWA and the low-ionization X-ray absorber
are not too dissimilar, tentatively suggesting a possible physical connection.
However, it is not possible to definitively link the DLWA and the 
low-ionization X-ray absorber in velocity space, as,
other than the Fe UTA and \ion{O}{7} edge, there are no strong X-ray absorption features $>$0.35 keV
endemic to only the low-ionization, and not the high-ionization,
X-ray absorber at the best-fit ionization parameter values.




Crenshaw \et\ (2001) note that the column density of the DLWA is $2 \times 10^{21}$ cm$^{-2}$,
very similar to that measured for both 
X-ray absorbers here. The outflow velocities relative to systemic
of DLWA lines (tens of km s$^{-1}$) are much lower than that of at
least the high-ionization X-ray absorber ($\sim 1500-2000$ km s$^{-1}$).  
Furthermore, it is likely that the DLWA has a high covering fraction (Crenshaw \et\ 2001).
We cannot rule out the possibility that the low-ionization X-ray absorber may be the inner
edge of the extended DLWA.
Moreover, it is conceivable that the outflowing, ionized X-ray absorbers
may supply the extended, 100-pc scale DLWA gas (at least along the line of sight), 
with the velocity
slowing and ionization parameter decreasing as the distance from the black hole increases.  
If the X-ray ionized absorbing gas is continuously feeding 
the UV absorber, we might expect the existence of an ``intermediate'' 
zone of absorbing material with outflow velocity  
between those of the observed X-ray and UV ionized absorbers. 
Confirmation of such a zone would strengthen the physical connection
between the X-ray and UV ionized absorbers. 
However, clarification of whether or not the X-ray absorbing outflow is
sustained or intermittent would be needed to determine the
exact physical and kinematic connection between the UV and X-ray absorbers.

The properties we have derived for the ionized X-ray absorbers are
consistent with the notion that dust may be swept up by an outflowing wind at radii 
$\gtrsim$ a few light-days (outside the BLR, where dust cannot survive). 
However, the current data cannot distinguish between a dusty and a dust-free
X-ray ionized absorber; as discussed in $\S$5, the upper limit on the neutral Fe L3
dust edge in the RGS spectrum corresponds to a 
hydrogren column density consistent with that 
implied by optical line absorption.


\subsection{The variable soft excess in NGC 3227} 
    
The origin of the soft excess emission in NGC 3227 is not immediately obvious.
The soft excess does not undergo the same obscuration by cold material as that 
suffered by the hard X-ray continuum.
However, the large discrepancy between $\Gamma_{\rm SX}$ and $\Gamma_{\rm HX}$
argues against both partial covering scenarios and against the 
soft X-rays being nuclear power-law emission scattered in an extended region.
We have modeled the soft excess in NGC 3227 using a blackbody component to represent
direct thermal emission from accretion disk, but such a model seems to be unphysical
given the apparent consistency of blackbody parameters across a wide range of Seyfert 
properties, including black hole mass (e.g., Gierli\'{n}ski \& Done 2004).
We have also modeled the soft excess as inverse Comptonization of seed photons, likely
thermal optical and UV photons from the accretion disk, by thermal electrons 
(Sunyaev \& Titarchuk 1980). Such a mechanism could conceivably operate in
the hot, ionized surface of the accretion disk (Hubeny \et\ 2001; Janiuk, 
Czerny \& Madejski 2001).


In Seyferts, soft excesses have been seen to vary on time scales of weeks, e.g., 
as seen in the narrow line Seyfert 1s Ark 564 and Ton S180 (Edelson \et\ 2002).
The soft excess in the quasar 3C~273 has also been known to vary on
time scales of $\sim$ a week (Kim 2001 and references therein).
Vaughan \et\ (2002) found the soft excess of the NLSy1 Ton S180
to vary rapidly, though in concert with the hard X-ray power law
component, leading to measurents of $F_{\rm var}$ roughly 
independent of energy band across the EPIC bandpass.
However, most other Seyferts' $F_{\rm var}$ spectra peak near 1--2 keV (e.g., Ar\'{e}valo \et\ 2008; 
Vaughan \& Fabian 2004). This is commonly explained with a variable
power-law component superimposed over a constant or relatively less variable hard component 
above $\sim$5 keV (likely the Compton reflection hump) and a constant or less variable
soft component below $\sim$1 keV (the soft excess). In NGC 3227, however, $F_{\rm var}$ increases
with decreasing energy below 1 keV (Figure 14) due to a soft excess that is strongly
variable in normalization.


The simultaneous increases in UV and soft X-ray continuum flux are not statistically significantly
correlated, so a direct UV--soft X-ray connection is tenuous at best.
However, if such a connection were real, and if both the UV and soft X-ray originate in the same 
location on the disk, a rapid change in illumination of the accretion disk could explain 
the simultaneous increases.

We can investigate if the observed UV trend can be attributed to
thermal reprocessing of the increasing soft X-ray flux trend
by the accretion disk.
Given the black hole mass, accretion rate, and assuming a standard thin
disk with an inner radius of 6 $R_{\rm g}$, 
90\% (50\%, 10\%) of the 260 nm continuum 
emission originates from within 60 (25, 10) $R_{\rm g}$.
The light crossing time across the diameter of the disk at
a 60 $R_{\rm g}$ radius is roughly 20 ks. 
In a thermal reprocessing scenario, the UV flux should track
the soft X-ray flux and be smeared on a time scale
$<$ 20 ks, assuming that the soft X-ray continuum emitting region is
located near the central disk plane.  
If the soft X-ray emitting region is located well off the plane,
the light travel time to the UV-emitting part of the disk
is increased (e.g., 30 ks travel time for a height of 100 $R_{\rm g}$.) 
The light curves displayed in Figure 1 are consistent
with this scenario, as any soft X-ray to UV lag is less than
several tens of ks.

Another possibility is that the soft X-ray continuum flux originates
in inverse Comptonization of UV seed photons from the disk; e.g., 
the scenario associated with the COMPST model.
However, the fact that the observed increase in the soft X-ray flux (40$\%$) 
is stronger than the observed increase in the UV flux (10$\%$)
would argue against all of the soft X-ray photons being passively-reprocessed
UV photons, unless only the soft X-ray photons were anisotropically beamed
along the line of sight, or unless the soft X-ray flux is responding to
a comparative increase in the UV continuum flux that
occurred less than a few tens of ks before the start of the 
{\it XMM-Newton} observation.  

In the inverse Comptonization scenario, and in the absence of thermal reprocessing,
the UV variability would have 
to be intrinsic to the disk. We can consider inwardly-propagating 
variations in the local mass accretion rate (e.g., Lyubarski 1997, 
Ar\'{e}valo \& Uttley 2006), specifically,
the case where the fluctuations travel on the viscous time scale.
To produce the 10$\%$ observed variation, such 
a propagating fluctuation would have to modify
20\% of the flux contained within a radius of 25 $R_{\rm g}$,
or 100\% of the flux contained within a radius of 10 $R_{\rm g}$.
Let us consider the former case: a 
fluctuation that propagates with an inward velocity that takes it from
a radius of 25 $R_{\rm g}$ to the innermost stable disk radius of
6 $R_{\rm g}$ over 100 ks, while modulating 
20\% of the flux within 25 $R_{\rm g}$, could
produce the observed UV variability. However, the required velocity,
$\sim$1/5 of the orbital velocity, is very high for a standard disk
(Shakura \& Sunyaev 1973): 
to be the viscous velocity for a standard disk, one would need
$(H/R)^2\alpha = 0.2$ ($H$, $R$, and $\alpha$ are the height, radius
and viscosity parameters, respectively), i.e., a very 
thick disk with a large viscosity parameter, which is probably 
not adequate for the optical-emitting portion of the disk.  
The UV variability is thus likely not associated with
accretion rate fluctuations traveling on viscous time scales.
The sound speed near 25 $R_{\rm g}$ (assuming $\alpha = 0.1$) is 
roughly two orders of magnitude slower than the orbital speed,
and sound waves are thus likely
inadequate to produce the required flux modification in 100 ks.



The rapidly variable soft excess is qualitatively similar to that
observed in 3C 120 with {\it Suzaku} by Kataoka \et\ (2006), who model
the variable soft excess in that object as the high-energy part of a 
synchrotron jet component. The rapid variability observed
in the soft X-ray and UV bands in NGC 3227, and the fact that the
soft excess undergoes less absorption by cold material 
compared to the hard X-ray continuum, may also be
consistent with existence of such a jet component,
in addition to and independent of the accretion disk corona 
(hard X-ray) emission. 
Unlike 3C 120, NGC 3227 is of course a radio-quiet AGN.
A $\sim$40 pc scale jet was detected in NGC 3227 by the VLA in 
a 1991 observation (Kukula \et\ 1995), though 10-100 pc scale jets are  
common in Seyferts (e.g., Gallimore \et\ 2006). 
Broad band spectral energy distribution (SED) modeling
of an accretion disk + jet is beyond the scope of this paper;
constraining such an SED component would be complicated by the presence of absorbing dust in 
the poorly-studied EUV band. However, such a component is plausible
if its peak in $\nu$$L_{\nu}$ space were near $10^{14-16}$ Hz 
(explained if the jet were young, e.g.,
$10^{\sim4-5}$ yrs), and the total observed synchrotron power were 
no more than $10^{41-42}$ erg s$^{-1}$. In this scenario, the observed  
UV continuum emission would be a mixture of slowly-varying "big blue bump"
emission from the accretion disk plus rapidly variable synchrotron emission
(which would contribute less than $\sim$50$\%$ to the total observed UV
continuum flux).


\subsection{The high energy emission components}

The photon index of the power-law component used to model the
hard X-ray continuum emission in both the
{\it XMM-Newton} and {\it RXTE} data 
is rather low ($\Gamma_{\rm HX} \sim 1.5-1.6$) compared to 
traditional or "canonical" values of $\sim 1.8-1.9$ for most broad-line
Seyfert 1 AGN.  However, similarly low values of 
$\Gamma_{\rm HX}$ have been reported previously for this object (e.g.,
George et al.\ 1998b). The presence of a cutoff near 100 keV, as implied by both
the {\it Swift}-BAT and the {\it RXTE}-HEXTE data,  
would suggest thermal Comptonization, but better
data in the 50--200 keV band and above are needed to confirm the 
cutoff and better constrain its energy.
The low-energy rollover to the hard X-ray continuum component was modeled above
as being due to cold absorption, but could also potentially be a signature of 
Comptonization, as it is possible for
Comptonized continuua to have a low-energy rollover e.g., below 1 keV
depending on parameters such as the distribution of input photon field
(e.g., Titarchuk 1994).

The 7.1 keV edge is likely due to reflection, as the absorbing
components modeled here predict an edge depth much too small 
to detect here. However, the strength of the Compton
reflection component $R$ is measured to be low ($\lesssim 0.5$),
suggesting that Compton-thick material exists in only a small
fraction of the sky as seen from the continuum source.
For example, the putative Compton-thick 
molecular torus invoked in standard
Seyfert 1/2 unification schemes (Urry \& Padovani 1995) could be
weak or small. Another possibility is that the 
optically-thick accretion disk could be truncated,
or the inner accretion disk could transition into
an optically-thin, radiatively inefficient flow, e.g.,
an ADAF or RIAF (Narayan \& Yi 1994, 1995; Blandford \& Begelman 1999; Narayan \et\ 2000).  
Such flows are frequently 
invoked when describing low accretion rate accreting black 
hole systems (e.g., low luminosity AGN). 
In addition, Liu \et\ (2007) forward a model for 
low accretion rate flows in which optically-thin coronal 
matter condenses into a cool, optically-thick inner disk, 
for $L_{\rm Bol}/L_{\rm Edd} = 0.1-2\%$ (assuming 
values for the viscosity parameter $\alpha \sim 0.1 - 0.4$).
Condensation radii of a few tens of $R_{\rm g}$ are plausible.

Wu \& Gu (2008) noted that for both black hole X-ray binary systems and AGN 
accreting above a critical "transition" value 
of $L_{\rm Bol}/L_{\rm Edd}$, $\Gamma_{\rm HX}$ correlates
with $L_{\rm Bol}/L_{\rm Edd}$. Below this transition,
$\Gamma_{\rm HX}$ and $L_{\rm Bol}/L_{\rm Edd}$ anti-correlate.
Wu \& Gu (2008) suggest that the transition between
a thin disk and an ADAF-type flow occurs near
$L_{\rm Bol}/L_{\rm Edd} = 1\%$ for black hole X-ray binary
systems and near 0.3$\%$ for AGN.
The corresponding value of $\Gamma_{\rm HX}$
at this transition point is 1.5, similar to that observed
in NGC 3227 ($L_{\rm Bol}/L_{\rm Edd} = 1\%$). This suggests that the accretion flow in
NGC 3227 may be consistent with either a thin disk, a ADAF-type flow,
or a transition to both.



\subsection{The Narrow Fe K$\alpha$ line at 6.4 keV}

The measured energy of the Fe K$\alpha$ emission line is consistent
with an origin in neutral material. The line is resolved in the pn spectrum;
the measured width $\sigma$ of
65$\pm$14 eV translates into a FWHM velocity $v_{\rm FWHM}$
of 7000$\pm$1500 km s$^{-1}$.
This velocity width is similar to that for the 
H$\beta$ line, 5500$\pm$500 km s$^{-1}$ (Wandel \et\ 1999), suggesting
an origin for the Fe K line in material commensurate with the BLR.

Assuming Keplerian motion, and assuming the velocity dispersion
is related to the FWHM velocity as $<$$v^2$$> = \frac{3}{4} v_{\rm FWHM}^2$
(Netzer \et\ 1990), assuming a black hole mass $M_{\rm BH}$ of
$4.22 \pm 2.14 \times 10^7 \Msun$ (Peterson \et\ 2004), we can use
$GM_{\rm BH} = r$$<$$v^2$$>$ to estimate the radius $r$ of the line-emitting
material. We find $r = 7.2^{+12.7}_{-4.9}$ light-days, equivalent to
$3000^{+5300}_{-2100}$ $R_{\rm g}$.
This value is consistent with the radius of $<$ 700 light-days
inferred from the very crude reverberation mapping discussed in $\S$6.2. 
It is also consistent with the BLR (H$\beta$) region radius of
$9^{+6}_{-8}$ light-days (Peterson \et\ 2004),
as well as with the $\sim$5--20 light-day
inner radius of the dust as reverberation-mapped by Suganuma \et\ (2006).

The Fe line-emitting material may be spatially extended in NGC 3227;
only the innermost regions have had time to respond to
continuum variations, which may get smoothed out by
the outer regions anyway. Alternatively, we note in Figure 16 that extrapolation of the 
$F_{2-10}$--$I_{\rm FeK\alpha}$ relation to zero continuum flux results in a non-zero offset.
That is, there may exist Fe-line emitting material which does not respond to the
continuum flux, and only some fraction of the total observed
Fe line flux responds to the continuum variations on $<$ 700 light-days.
Either of these two ideas may explain why
$F_{\rm var}$ for the 2--10 keV continuum flux is 
larger than that for the Fe line.

The observed line $EW$ is 91$\pm$10 eV.  
The column density of the gas required to produce such an emission
line must be greater than $10^{\sim 22}$ cm$^{-2}$, 
otherwise the optical depth would be insufficient to produce
such a prominent line. However, the lack of an obvious 
Compton shoulder (the emission near 6.0 keV
was not identified as such) would argue against the bulk of the line
photons originating in Compton-thick material.

Let us first assume an origin in optically-thin gas that completely surrounds
a single, isotropically-emitting continuum source (covering fraction 
$f_{\rm c}$ = 1)
and is uniform in column density. 
We can relate $EW$ to $N_{\rm H}$ using the following equation,
taken from e.g., Markowitz \et\ (2007):
\begin{equation}
EW_{\rm calc} = f_{\rm c} \omega f_{\rm K\alpha} A \frac{\int^{\infty}_{E_{\rm K
 edge}}P(E) \sigma_{\rm ph}(E) N_{\rm H} dE}{P(E_{\rm line})}
\end{equation}
$\omega$ is the fluorescent yield: the value for Fe, 0.34, was taken from 
Kallman \et\ (2004). $f_{\rm K\alpha}$ is the fraction of 
photons that go into the K$\alpha$ line
as opposed to the K$\beta$ line; this is 0.89 for \ion{Fe}{1}.
$A$ is the number abundance relative to hydrogen;
solar abundances, using Lodders (2003), were used.
$P(E)$ is the spectrum of the illuminating continuum at energy $E$;
$E_{\rm line}$ is the K$\alpha$ emission line energy.
$\sigma_{\rm ph}(E)$ is the photo-ionization cross section 
assuming absorption by K-shell electrons only; all cross sections were taken
from Veigele (1973\footnote{http://www.pa.uky.edu/$\sim$verner/photo.html}).
The observed Fe K$\alpha$ $EW$ can thus be produced from material with
$N_{\rm H} = 1.4 \pm 0.2 \times 10^{23}$ cm$^{-2}$.
Similarly, the observed Ni K$\alpha$ line $EW$ of 13$^{+13}_{-10}$ eV
indicates $N_{\rm H} = 3.9^{+3.9}_{-2.9} \times 10^{23}$ cm$^{-2}$.

If the absorbers are instead clumpy and lie out of the line of sight, we
can use Eqn.\ 1 of Wozniak \et\ (1998), which gives the expected Fe K$\alpha$
line intensity for a cloud with a column $\gtrsim 10^{23}$ cm$^{-2}$ 
lying off the line of sight and subtending a fraction $\Omega/4\pi$ 
of the sky as seen from a single, isotropically-emitting continuum
source. Assuming solar abundances and given the observed 
Fe K$\alpha$ line intensity, for values of $\Omega/4\pi$ =
0.1 (0.3), a column density of $1.5 \times 10^{23}$ 
($5 \times 10^{22}$) cm$^{-2}$.

In either case, the derived column density is higher than
that of the hard X-ray absorbing material in the SXPL
and COMPST model fits to the {\it XMM-Newton} EPIC data. 
It is, however, closer to the column density
of the obscuring cloud during the 2000-1 obscuring event (Lamer \et\ 2003).
Based on the duration of the obscuring event and the implied
number density, that cloud was inferred to be located in the BLR.
The similarities in column density and inferred location
suggest a connection between the Fe line-emitting material
and the obscuring cloud of 2000-1.


\subsection{Emission at 6.0 keV}

The Fe K emission profiles in both the pn and MOS 1+2 spectra 
suggest emission redward of the Fe K$\alpha$ core. However, the present
data cannot distinguish between a model wherein the emission
near 6.0 keV is due to a narrow emission line or is instead the
red wing of a relativistically broadened Fe K$\alpha$ line with
an $EW$ near 80 eV. 

\subsubsection{A narrow emission feature?}

We first discuss the emission in the context of modeling it as a narrow
feature. In the pn spectrum, the line width $\sigma$ was $<$200 eV; the 
$EW$ was $21^{+19}_{-9}$ eV. High-resolution spectroscopy has yielded
evidence for similar narrow emission features in
roughly 16 Seyferts to date (see Vaughan \& Uttley 2008 for a review).
These features are commonly interpreted as red- or blue-shifted Fe K features
associated with "hot-spots," localized areas of the accretion disk
illuminated by a localized flare just above the disc (possibly due 
to magnetic reconnection), rather than a central 
illuminator or an extended corona (e.g., Dov\u{c}iak et al.\ 2004, 
Goosmann et al.\ 2007).

Vaughan \& Uttley (2008) point out that these narrow emission
features are rarely reported at more than 3.0$\sigma$ confidence.
Consideration of selection bias and publication bias 
brings skepticism to the validity of these features and
raises the possibility that a non-trivial fraction of 
these published lines may be consistent with being due to photon noise.
In the case of the pn spectrum of NGC 3227,
the Monte Carlo simulations we performed suggested that
the line was detected at $>$99.9$\%$ confidence (i.e., $<$0.1$\%$
likelihood that it is due to photon noise). However,
we caution that the Monte Carlo procedure yields an estimate of the 
detection significance independently of the existence of
other high-resolution spectra of NGC 3227 and other Seyferts.  
On the other hand, the marginal detection of the emission feature near 6.1 keV
in the MOS 1+2 spectrum, simultaneously to the pn data, reduces the
likelihood that the feature is due to photon noise. In addition,
the value of $\vert\Delta\chi^2\vert$ associated with modeling the
line in the pn spectrum is among the largest reported for 
narrow emission lines (See Table 1 of Vaughan \& Uttley 2008).

If the feature near 6.0 keV is indeed emission from
a "hot spot" on the accretion disk, we can estimate its radial distance from the black hole.
For instance, we can assume that the observed redshifting is due solely 
to Doppler shifting associated with Keplerian motion of the receding side 
of the disk. We use the 3$\sigma$ energy range of 5.8 to 6.3 keV, 
corresponding to radial velocities along the line of sight 
$v_{\rm r}$ of 0.02--0.09$c$ (assuming an origin in neutral Fe).
Assuming the disk is inclined by $i$ = 30$\degr$ (45$\degr$) with respect to
the plane of the sky, using $M_{\rm BH} = 4.2 \times 10^7 \Msun$ 
(Peterson \et\ 2004), and ignoring gravitational redshifting, this 
corresponds to Keplerian motion at radii of roughly 
12--430 (25--800) $R_{\rm Sch}$.

If the flare is co-rotating with the disk, then a minimum radius 
independent of $i$ is defined by the fact that the line energy is
consistent with being constant and redward of 6.4 keV for the 100 ks
duration of the observation. Any orbital period $<$100 ks is excluded;
otherwise the number of photons $>$6.4 keV would be greater than 
the number of photons $<$6.4 keV. Radii $<$ 9 $R_{\rm Sch}$ are 
thus excluded.

An alternate possibility is that the line-emitting material is
in free-fall, with no velocity component transverse to 
our line of sight. Yet another possibility is that the observed redshifting
is purely due to energy losses of photons escaping from 
the vicinity of the black hole; 
assuming for simplicity emission from neutral Fe originating in gas lying
along the line of sight to the black hole, with all
velocity transverse to line of sight, radii of 
$\sim 6-30 R_{\rm Sch}$ are plausible.

Another speculative possibility is that the line-emitting material could be
associated with the base of the jet, forming above the accretion disk,
as opposed to being a part of the accretion disk.
Close to the rotation axis in AGN,
magnetic fields are likely twisted by the differential 
rotation of the disk, serving to launch and collimate jets 
(e.g., Blandford \& Payne 1982).
For instance, Marscher \et\ (2008) presented observational evidence for
the jet in the radio-loud AGN BL Lac to follow a spiral flow as it is 
accelerated through a zone containing a helical magnetic
field. The 6.0 keV line in NGC 3227 may be emission from Fe in a blob
of material, with a column density $N_{\rm H}$ of 10$^{\sim 21-22}$ cm$^{-2}$,
which has been caught up in the launching and formation
of the weak jet. If the material is close to the axis, 
above the disk, and moving with a velocity with a large azimuthal 
component, then an observed redshift is possible unless the 
rotation axis lies very close to the line of sight.
However, the weakness of Seyferts' jets may possibly be attributed to
weaker acceleration and/or collimation mechanisms compared to
those in blazars, reducing the likelihood that 
such azimuthal velocities can exist in Seyferts.

\subsubsection{A relativistically broadened diskline component?}

Evidence for relativistically broadened Fe line 
"diskline" profiles in a significant fraction of Seyferts
has been accumulating since the days of {\it ASCA}
(Tanaka \et\ 1995, Fabian \et\ 2000).
In recent years, {\it XMM-Newton} and {\it Chandra}-HETGS
observations have demonstrated that a narrow core at 6.4 keV 
is ubiquitous; accurately modeling that component, as well as
absorption due to ionized gas in the line of sight, is critical
for accurate measurement of the flux and profile of 
any broad Fe line present. 

In NGC 3227, our best-fit "DL"
model incorporates a narrow Fe K$\alpha$ core,
and suggests a relatively weak broad line, with an $EW$ of
81$^{+42}_{-30}$ eV.  We constrained the inner radius to be $< 22 R_{\rm g}$ and the inclination to be
$<$25$\degr$ (though these values were derived with the radial
emissivity parameter fixed at --3).

Assuming solar abundances, the abundances of Lodders (2003),
and given the measured strength of the Compton reflection component 
$R = 0.40 \pm 0.07$, the predicted Fe line $EW$ 
(George \& Fabian 1991)\footnote{George \& Fabian 
1991 predict $R$ = $EW$/150 eV, using
the abundances of Morrison \& McCammon (1983), which assumed
$3.3 \times 10^{-5}$ Fe atoms per H atom. Lodders (2003) sets
the Fe abundance to $2.95 \times 10^{-5}$ Fe atoms per H atom.} 
is 54$\pm$9 eV, consistent with the modeled $EW$ of the observed 
broad Fe line.  The low value of the broad line $EW$ may also suggest
that the optically-thick disk is not spatially 
extended, again consistent with the notion that
the nucleus of NGC 3227 could harbor an ADAF or RIAF,
and the upper limit on the inner radius of 22 $R_{\rm g}$ is 
potentially consistent with the model of Liu \et\ (2007) containing
a small, optically-thin disk.


\section{Conclusions}

We have observed the nucleus of the Seyfert 1.5 AGN NGC 3227
with {\it XMM-Newton} in December 2006 for almost 100 ks.
We have used EPIC and RGS spectra to study the Fe K bandpass,
ionized absorbers, and 0.2--10 keV continuum in detail. We also present
X-ray continuum light curves and a UV continuum light curve obtained with the 
Optical Monitor. We have combined these data with archival {\it RXTE}-PCA and HEXTE
monitoring, plus {\it Swift}-BAT monitoring data, to constrain the level
Compton reflection, study the high energy continuum up to 200 keV,
and track the flux behavior of the Fe K line on time scales of weeks to years.
Our main results are summarized as follows:

The EPIC pn spectrum shows the prominent narrow Fe K$\alpha$ emission line
to be consistent with an origin in neutral Fe (Gaussian energy centroid
6.403 $\pm$ 0.009 keV). Its intensity and 
$EW$ are $3.5 \pm 0.4 \times 10^{-5}$ ph cm$^{-2}$ s$^{-1}$ and 91$\pm$10 eV,
respectively, consistent with an origin in material with a column density
$10^{\sim22.5-23}$ cm$^{-2}$. The line is resolved in the pn spectrum, with
a FWHM velocity 7000$\pm$1500 km s$^{-1}$. Assuming purely Keplerian motion,
we estimate the radius of the line-emitting material to be 
$7.2^{+12.7}_{-4.9}$ light-days.
The FWHM velocity and estimated radius are consistent with the
BLR (as mapped by H$\beta$ emission)
as well as with the $\sim$5--20 light-day
inner radius of the dust as reverberation-mapped by Suganuma \et\ (2006).

Time-resolved spectral fitting to the 1999--2005 {\it RXTE}-PCA monitoring
data reveal tentative evidence for a significant fraction of the
Fe K line photons to track variations seen in the continuum, with a 
light travel time delay which is not tightly constrained, but is
$<$ 700 days. We thus rule out the 
possibility that the bulk of the (variable) Fe line photons
originate at distances of 1--2 pc and more, e.g., in a multi-pc-scale molecular torus.
However, more intensive sampling on time scales of days to weeks is required
to beter constrain the lower limit on the continuum--line lag.

Emission near 6.0 keV is detected in both the pn and MOS 1+2 spectra.
The fact that it is detected in both instruments and the relatively
large associated value of $\Delta\chi^2$ both argue against this feature being
an artifact due to photon noise.
It is inconsistent with being a Compton shoulder to the 6.4 keV
Fe K$\alpha$ line.
We modeled the emission using a narrow Gaussian component, with
energy centroid (in the pn spectrum) 6.04$^{+0.18}_{-0.04}$ keV,
width $\sigma < 200$ eV, intensity 
$9^{+8}_{-4} \times 10^{-6}$ ph cm $^{-2}$ s$^{-1}$ and 
$EW$ of 21$^{+19}_{-9}$ eV. A possible origin is a "hot-spot" in the 
accretion disk, originating at radii of tens to hundreds of
$R_{\rm g}$. The emission feature is modeled equally well as the red wing
to a relativistically broadened "diskline" component, with
inner radius $<$22 $R_{\rm g}$, inclination $<$25$\degr$, 
and $EW$ = 81$^{+42}_{-30}$ eV.

Broadband (0.2-10 keV) EPIC spectral modeling reveals a strong soft
excess dominating below 1 keV. In the pn spectrum, the soft excess
is fit well by a steep power law, with $\Gamma_{\rm SX}$ $\gtrsim$ 3.
A blackbody component and a low-temperature Comptonization component  
each fit the data well, though the physical plausibility these latter two
components in Seyferts is not generally accepted.
The normalization of the soft excess increases by about 20$\%$ during the
first $\sim$20 ks of the {\it XMM-Newton} observation, and by 
$\sim$40$\%$ over $\sim50$ ks. The soft X-ray band is more strongly
variable than the hard X-ray band on time scales of tens of ks:
unlike the $F_{\rm var}$ spectra of many other Seyferts,
the $F_{\rm var}$ spectrum of NGC 3227 below 1-2 keV continues to increase as photon
energy decreases. Such relatively rapid variability in the
soft excess, independent of rapid variability in the hard X-ray band
is very rare for Seyferts.

The OM shows the UV continuum flux (near 260 nm) to
increase by 10$\%$ during the {\it XMM-Newton} observation.
Such $\lesssim$1 day variability is relatively strong compared to
similar-duration UV continuum light curves obtained with the OM
(Smith \& Vaughan 2007).

We present the highest-quality gratings spectrum obtained for NGC 3227 to date.
In the RGS spectrum, we model the ionized absorption using two zones.
The higher-ionization zone has log$\xi_{\rm hi}$ = $2.90^{+0.21}_{-0.26}$,
and an ouflow velocity relative to systemic of  --(2060$^{+240}_{-170}$) km s$^{-1}$.
This indicates a minimum radial distance from the black hole of 40 light-days, 
placing it outside the BLR radius. 
The lower-ionization zone has log$\xi_{\rm lo}$ = $1.21^{+0.18}_{-0.08}$ 
(corresponding to a factor of roughly 3--10 higher than the 
dusty lukewarm absorber modeled by Kraemer \et\ 2000 and 
Crenshaw \et\ 2001). Its main signature is a Fe M shell UTA near 740--780 eV.
The best estimate for its outflow velocity relative to systemic is
--(420$^{+430}_{-190}$) km s$^{-1}$.
We find no evidence for non-solar abundances.
Both absorbing zones have column densities $N_{\rm H,WA}$
near $ 1-2 \times 10^{21}$ cm$^{-2}$. $N_{\rm H,WA}$
and the column densities of local, cold absorption measured from the EPIC spectra
are too low to be directly associated with the 
obscuring BLR cloud in 2000-1 studied by Lamer \et\ (2003).
However $N_{\rm H,WA}$ is similar to that of the 100-pc scale DLWA,
consistent with the notion that the outflowing X-ray absorbing material
may supply the UV-absorbing DLWA material.

In the RGS spectrum, we detect five narrow emission lines detected at the systemic redshift; we
identify these lines as being due to \ion{N}{6}  {\it (r)}, \ion{N}{7} , and
the {\it (f)}, {\it (i)}  and {\it (r)} lines of \ion{O}{8}.  
No emission due to \ion{Fe}{17} L  line (3d--2p) at 826 eV
(a signature of collisionally-ionized plasma) is detected
(upper limit of 1 eV).
No strong RRC features due to H- or He-like ions
(signatures of photo-ionized plasma) are detected
(upper limits range from 5 to 50 eV).

We presented the total spectrum derived from 
{\it RXTE}-PCA and HEXTE archival monitoring data taken in 1996 and from 1999--2005,
and combined them with a 4-channel Swift-BAT spectrum
from the 9-month survey data (obtained in 2005).
The HEXTE and BAT spectral data reported here represent the first detailed
spectrum of NGC 3227 above $\sim$ 20 keV and up to almost 200 keV.
The hard X-ray continuum photon index in both the {\it RXTE}+BAT
spectra and {\it XMM-Newton} EPIC spectra is rather flat, $\Gamma_{\rm HX} = 1.5-1.6$.
The strength of the Compton reflection hump is rather low,
$R \lesssim 0.5$, consistent with the $EW$ of the diskline component
modeled in the pn spectrum (assuming solar abundances).
We also find evidence for a high-energy continuum cutoff at 90$\pm$20 keV.

The low values of $\Gamma_{\rm HX}$ and $R$, combined with the
low value of $L_{\rm Bol}/L_{\rm Edd} \sim 1\%$, are consistent with the
notion that NGC 3227 may harbor an optically-thin, radiatively-inefficient flow
such as an ADAF, in addition to or instead of
a standard geometrically-thin, radiatively-efficient disk.



\acknowledgements 
A.M.\ thanks M.\ Elvis, A.\ Marscher and D.\ Evans for
helpful suggestions. This work is based on an observation 
obtained with {\it XMM-Newton}, an ESA science mission, and
has made use of HEASARC online services, supported by
NASA/GSFC, the NASA/IPAC Extragalactic Database,
operated by JPL/California Institute of Technology under
contract with NASA, and the NIST Atomic Spectra Database.



\clearpage

\begin{deluxetable}{llc}
\tabletypesize{\footnotesize}
\tablewidth{5.0in}
\tablenum{1}
\tablecaption{Fe K bandpass components  
in the time-averaged EPIC pn spectrum\label{tab1}}
\tablehead{
\colhead{Component} & \colhead{Parameter} & \colhead{Value}}
\startdata 
             &  $\chi^2$/$dof$     & 983.2/990  \\
Fe K$\alpha$ emission line &  Energy (keV) &   6.403 $\pm$ 0.009 \\
                  &  Width $\sigma$ (eV)  & 65 $\pm$ 14  \\ 
    & Intensity (ph cm$^{-2}$ s$^{-1}$)&    $3.5 \pm 0.4 \times 10^{-5}$ \\
    & $EW$ (eV) &  91 $\pm$ 10  \\
Ni K$\alpha$ emission line &  Energy (keV) &   $7.41^{+0.08}_{-0.07}$ \\
    & Intensity (ph cm$^{-2}$ s$^{-1}$) &    $4^{+4}_{-3} \times 10^{-6}$ \\
    & $EW$ (eV)   &  $13^{+13}_{-10}$ \\
6.04 keV emission line   & Energy (keV) &    $6.04^{+0.20}_{-0.06}$ \\
                & Width $\sigma$ (eV) &  $<$ 200 \\
    & Intensity (ph cm$^{-2}$ s$^{-1}$) &   $9.1 \pm 2.5 \times 10^{-6}$ \\
    & $EW$ (eV) &  22 $\pm$ 6  \\
Fe K Edge  &   Energy (keV)  & 7.11 (fixed) \\
    &  Optical Depth $\tau$  & 0.05 $\pm$ 0.03  \\
\enddata
\tablecomments{Parameters for the best-fitting 
``GA'' model (wherein the emission near 6.0 keV is modeled with a narrow Gaussian component),
fit to $>$4 keV data.
The model includes a Fe K$\beta$ line, with energy fixed at 7.056 keV
(rest frame), width $\sigma$ tied to that of the K$\alpha$ 
line, and intensity equal to 0.13 times that of the K$\alpha$ line. 
The width $\sigma$ for the Ni K$\alpha$ line was also
tied to that for the Fe K$\alpha$ line.
Equivalent widths $EW$ were determined relative to a locally-fit continuum. }
\end{deluxetable}


\begin{deluxetable}{llccc}
\tabletypesize{\footnotesize}
\tablewidth{6.5in}
\tablenum{2}
\tablecaption{Model fits to the 0.2--10 keV time-averaged EPIC pn spectrum\label{tab2}}
\tablehead{
\colhead{Component} & \colhead{Parameter} & \colhead{SXPL Model} & \colhead{BB Model} & \colhead{COMPST Model} }
\startdata
      &  $\chi^2$/$dof$  &   1866.9/1724 &   1870.4/1724 &  1863.4/1723 \\
Local cold absorption & $N_{\rm H,local}$  (cm$^{-2}$)      &   $8.7^{+0.6}_{-0.5} \times 10^{20}$ &  $5.1^{+1.2}_{-0.2} \times 10^{20}$ &   $7.7^{+0.7}_{-0.3} \times 10^{20}$ \\
Hard X-ray power-law  & $\Gamma_{\rm HX}$  &  1.57 $\pm$ 0.02     &        1.61$^{+0.01}_{-0.02}$  &     1.65$^{+0.02}_{-0.03}$ \\
                      & Norm.\ (1 keV)  &    $6.7^{+0.3}_{-0.2} \times 10^{-3}$ &  $7.4 \pm 0.1 \times 10^{-3}$  & $8.1^{+0.2}_{-0.5} \times 10^{-3}$  \\
Hard X-ray Absorption &  $N_{\rm H,HX}$ (cm$^{-2}$) &     $2.9^{+0.3}_{-0.8} \times 10^{21}$  &            $< 6.1 \times 10^{20}$  &  $4.2^{+0.3}_{-1.1} \times 10^{21}$ \\
Low-ioniz.\ X-ray Warm Abs.\ & $N_{\rm H,lo}$ (cm$^{-2}$) &     $1.0^{+0.3}_{-0.1} \times 10^{21}$  &   $7.9^{+2.7}_{-0.7} \times 10^{20}$ &    $9.3^{+2.1}_{-1.9} \times 10^{20}$ \\
                           & log$\xi_{\rm lo}$ (erg cm s$^{-1}$)   &  $1.45^{+0.16}_{-0.07}$  &   1.38 $\pm$ 0.19  &        $1.43^{+0.20}_{-0.43}$ \\
High-ioniz.\ X-ray Warm Abs.\ &  $N_{\rm H,hi}$ (cm$^{-2}$) &     $1.8^{+1.2}_{-0.6} \times 10^{21}$ &   $1.8^{+1.3}_{-0.4}\times 10^{21}$ &  $1.5^{+1.3}_{-0.6} \times 10^{21}$ \\
                           & log$\xi_{\rm hi}$ (erg cm s$^{-1}$)   &  $2.93^{+0.15}_{-0.09}$ & 2.91$^{+0.15}_{-0.07}$  &  $2.89^{+0.14}_{-0.20}$ \\ 
Soft X-ray power-law    &  $\Gamma_{\rm SX}$  &  $3.35^{+0.27}_{-0.10}$ &    &     \\
                        &  Norm.\ (1 keV)  & $4.0^{+0.2}_{-0.7} \times 10^{-3}$  &  & \\ 
Blackbody              & Temperature $k_{\rm B}T$ (eV)     &      & $83^{+1}_{-4}$ &    \\
                       & Norm.\                             &    &   $1.72^{+0.37}_{-0.11} \times 10^{-4}$  &  \\
Comptonized component   &  Temperature $k_{\rm B}T$ (keV)   &     &         &   $0.35^{+0.02}_{-0.03}$ \\
                        & Optical Depth $\tau$              &     &         &  24$^{+2}_{-4}$ \\
                                    & Norm.\                            &     &         &  $4.41^{+0.22}_{-0.33} \times 10^{-3}$ \\    
\ion{O}{7} emission line   &  Energy (keV)      & $0.58 \pm 0.01$      & $0.58^{+0.02}_{-0.01}$      & $0.58 \pm 0.01$         \\
                            & Intensity (ph cm$^{-2}$s $^{-1}$) &   $2.4^{+0.5}_{-0.4} \times 10^{-4}$ & $1.2^{+0.6}_{-0.5} \times 10^{-4}$  &   $2.1 \pm 0.4 \times 10^{-4}$ \\
\enddata
\tablecomments{SXPL, 
BB and COMPST refer to the models wherein the soft excess is modeled with
a steep power-law component, a blackbody component, and a Comptonized component
({\sc CompST}), respectively. The units of normalization of the power-law
components are  ph keV$^{-1}$ cm$^{-2}$ s$^{-1}$ at 1 keV.
The blackbody normalization is $L_{39}D_{10}^{-2}$, where $L_{39}$ is the source luminosity in 
units of 10$^{39}$ erg s$^{-1}$ and $D_{10}$ is the distance to 
the source in units of 10 kpc. 
See the {\sc xspec} user manual for units of the {\sc CompST} component normalization.
The reader is reminded that the BB and COMPST models
are not generally accepted as physically plausible descriptions 
of soft excesses in Seyferts.
The uncertainties listed on the intensity of the \ion{O}{7} emission line
are statistical only and do not
include likely systematic uncertainties due
to blending with narrow absorption lines.
The width $\sigma$ of the \ion{O}{7} emission line was kept fixed at 0.5 eV.}
\end{deluxetable}


\begin{deluxetable}{lcccc}
\tabletypesize{\footnotesize}
\tablewidth{5.0in}
\tablenum{3}
\tablecaption{Emission lines fit to the RGS spectrum\label{tab3}}
\tablehead{
\colhead{Line}           & \colhead{Energy} & \colhead{Intensity}               & \colhead{}              &  \colhead{$F$-test} \\
\colhead{Identification} & \colhead{(eV)}   & \colhead{(ph cm$^{-2}$ s$^{-1}$)} & \colhead{$\Delta\chi^2$} &  \colhead{Probability} }
\startdata
\ion{N}{6}    {\it (r)}        &  431 $\pm$ 1        &   $9.9^{+6.7}_{-5.0} \times 10^{-5}$      &    --12.0  & 0.042 \\
\ion{N}{7}                   &    $499^{+1}_{-2}$    &   $8.2^{+6.0}_{-5.2} \times 10^{-5}$      &   --10.4  & 0.064  \\
\ion{O}{7}  {\it (f)}        &    $561^{+1}_{-2}$    &   $1.33^{+0.13}_{-0.75} \times 10^{-4}$      & --25.2  & $1.4 \times 10^{-3}$  \\
\ion{O}{7}  {\it (i)}        &    567 $\pm$ 1       &   $9.5^{+3.2}_{-3.5} \times 10^{-5}$      & --18.5  & $7.8 \times 10^{-3}$    \\
\ion{O}{7}  {\it (r)}        &   $572^{+1}_{-2}$    &   $4.9^{+12.3}_{-2.9} \times 10^{-5}$      & --7.9 & 0.12      \\
\enddata
\tablecomments{Best-fit line emission parameters using Gaussian components.
All line widths $\sigma$ were kept fixed at 0.5 eV.
The uncertainties listed on the intensities (Col.\ [3])
are statistical only and do not include likely systematic uncertainties due
to blending with narrow absorption lines.
The $F$-test probability values $P$ (Col.\ [5]) denote the
changes that the null hypothesis (line not included in model) is valid;
the line detection significance is 1 -- $P$.}
\end{deluxetable}


\begin{deluxetable}{llc}
\tabletypesize{\footnotesize}
\tablewidth{5.0in}
\tablenum{4}
\tablecaption{Best-fit model parameters for the RGS spectrum\label{tab4}}
\tablehead{
\colhead{Component} & \colhead{Parameter} & \colhead{Value}}
\startdata 
             &  $\chi^2$/$dof$     & 785.7/417   \\
Local cold absorption & $N_{\rm H,local}$  (cm$^{-2}$)      &   $10.5^{+1.2}_{-1.9} \times 10^{20}$ \\ 
Soft X-ray power-law    &  $\Gamma_{\rm SX}$  &    3.00$\pm$0.25 \\
                         &   Norm.\ (1 keV)  &   $6.2^{+0.3}_{-0.7} \times 10^{-3}$ \\    
Low-ioniz.\ X-ray Warm Abs.\ & $N_{\rm H,lo}$ (cm$^{-2}$) &     $1.1^{+0.1}_{-0.2} \times 10^{21}$ \\
                             & log$\xi_{\rm lo}$ (erg cm s$^{-1}$)   & $1.21^{+0.18}_{-0.08}$  \\
                             & $z_{\rm lo}$ (absolute)          &  $+0.00246^{+0.00144}_{-0.00064}$   \\
                             &  $z_{\rm lo}$ (rel.\ to systemic) &  $-0.00140^{+0.00144}_{-0.00064}$   \\
High-ioniz.\ X-ray Warm Abs.\ & $N_{\rm H,hi}$ (cm$^{-2}$) &     $2.4^{+2.0}_{-1.2} \times 10^{21}$ \\
                             & log$\xi_{\rm hi}$ (erg cm s$^{-1}$)   & $2.90^{+0.21}_{-0.26}$ \\
                             & $z_{\rm hi}$ (absolute)          &  $-0.00302^{+0.00057}_{-0.00080}$   \\
                             &  $z_{\rm hi}$ (rel.\ to systemic) &  $-0.00688^{+0.00057}_{-0.00080}$   \\
\enddata
\tablecomments{Results are shown for the best-fit soft X-ray power-law model, including
five emission lines due to He-like O and H- and He-like N; see Table~3
for emission line parameters. The units of normalization of the power-law
components are  ph keV$^{-1}$ cm$^{-2}$ s$^{-1}$ at 1 keV.
See text for details regarding uncertainty range in $z$.
The model also included a hard X-ray power-law component 
absorbed by a column $N_{\rm H,HX} \sim 8 \times 10^{21}$ cm$^{-2}$ and
emerging only above $\sim$ 1.5 keV.}
\end{deluxetable}


\begin{figure}
\epsscale{0.60}
\plotone{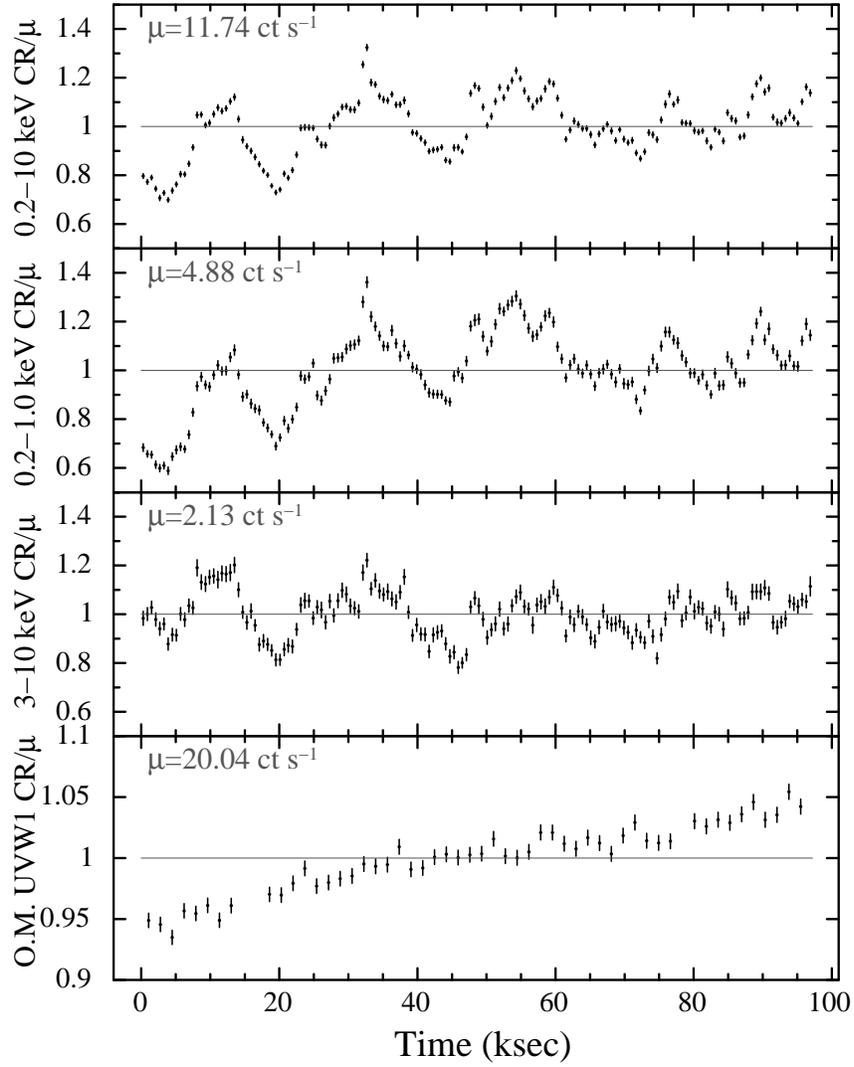}
\caption{The top three panels show the EPIC-pn count rate light curves
for the 0.2--10, 0.2--1, and 3--10 keV bandpasses, binned to 600 s. The bottom
panel shows the OM UVW1 light curve, binned to 1400 s.}
\end{figure}

\begin{figure}
\epsscale{0.75}
\plotone{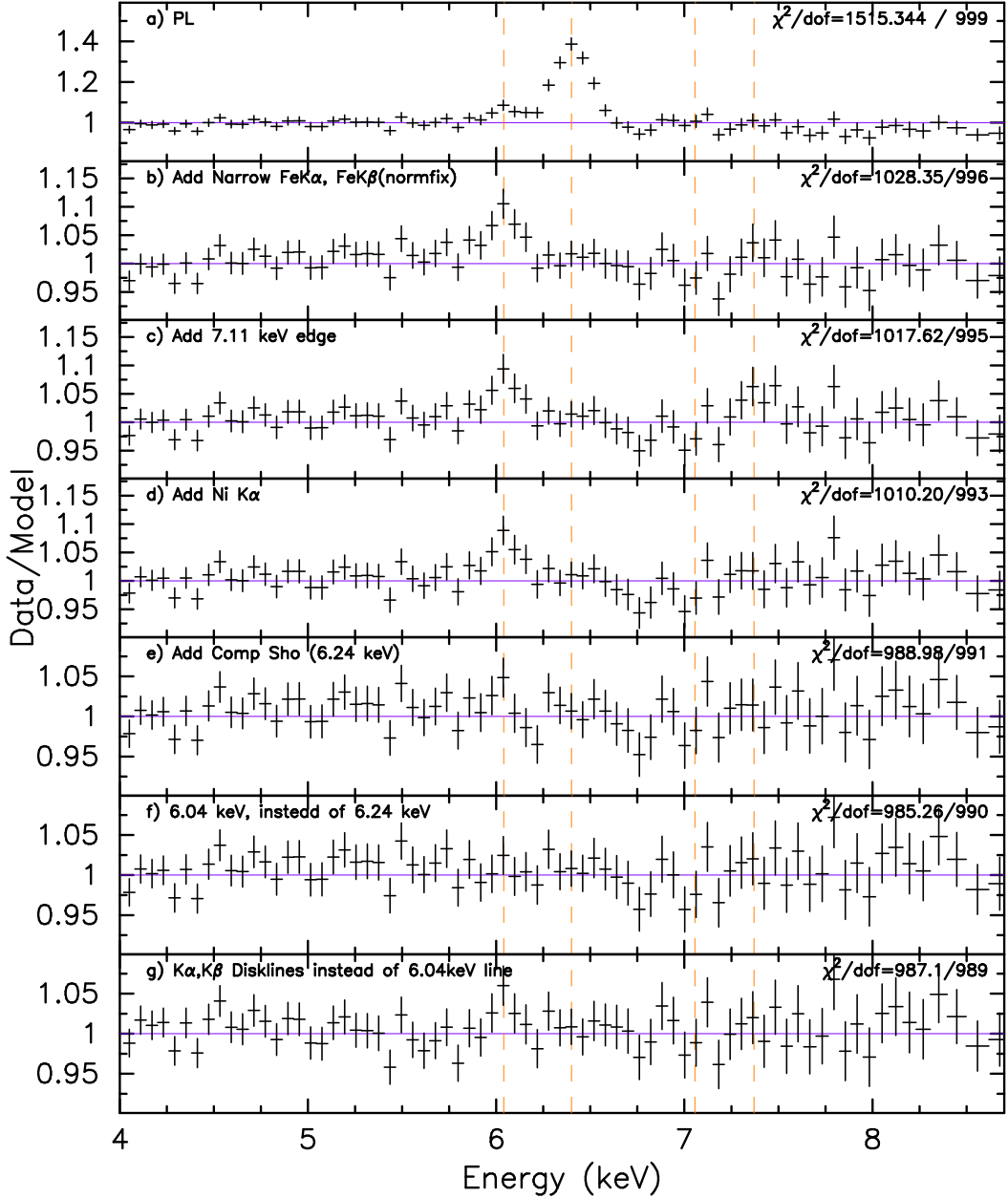}
\caption{Data/model residuals to spectral fits to the Fe K bandpass 
of the EPIC-pn spectrum. Data are rebinned by a factor of 12. Vertical dashed lines
denote energies of 6.04, 6.40, 7.06, and 7.37 keV. 
Panel {\it a)} shows residuals to a simple power-law model.
In panel {\it b)}, the Fe K$\alpha$ and Fe K$\beta$ emission lines have been
modeled. The 7.11 keV edge and Ni K$\alpha$ line
have been modeled in panels {\it c)} and {\it d)}, respectively.
In panel {\it e)}, a Gaussian with an energy centroid of 6.24 keV has 
been added. However, as shown in panel {\it f)}, the best-fit ``GA'' model,
with a Gaussian with an 
energy centroid at 6.04 keV, does a superior job in modeling the remaining
$\sim$6 keV residuals. The small dip near 6.8 keV is not
significant and is likely an artifact of fitting. 
In panel {\it g)}, the best-fit ``DL'' model,
the 6.04 keV line has been
removed and replaced with relativistically broadened Fe K$\alpha$ and
Fe K$\beta$ diskline profiles; the narrow Gaussian at 6.04 keV
(panel {\it f)}) seems to model the residuals better. }
\end{figure}


\begin{figure}
\epsscale{0.50}
\plotone{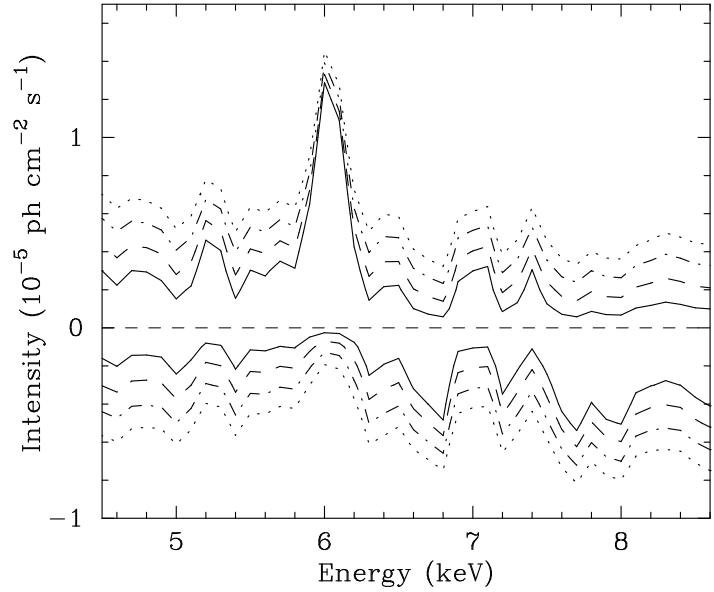}
\caption{Contour plots showing the results of applying a
``sliding Gaussian'' (with width $\sigma$ fixed at 10 eV)
to a model to the EPIC-pn data
consisting of a power-law plus a narrow Gaussian 
at 6.40 keV to model Fe K$\alpha$ emission.
1-, 2-, 3- and 4-$\sigma$
confidence levels for two interesting parameters are denoted by solid, dashed, dot-dashed, and dotted lines,
respectively.
Residuals near 6.0 keV are clear; additional residuals
at 7.0 and 7.4 keV (above the Fe K edge at 7.11 keV)
are investigated further in $\S$3.}
\end{figure}

\begin{figure}
\epsscale{0.50}
\plotone{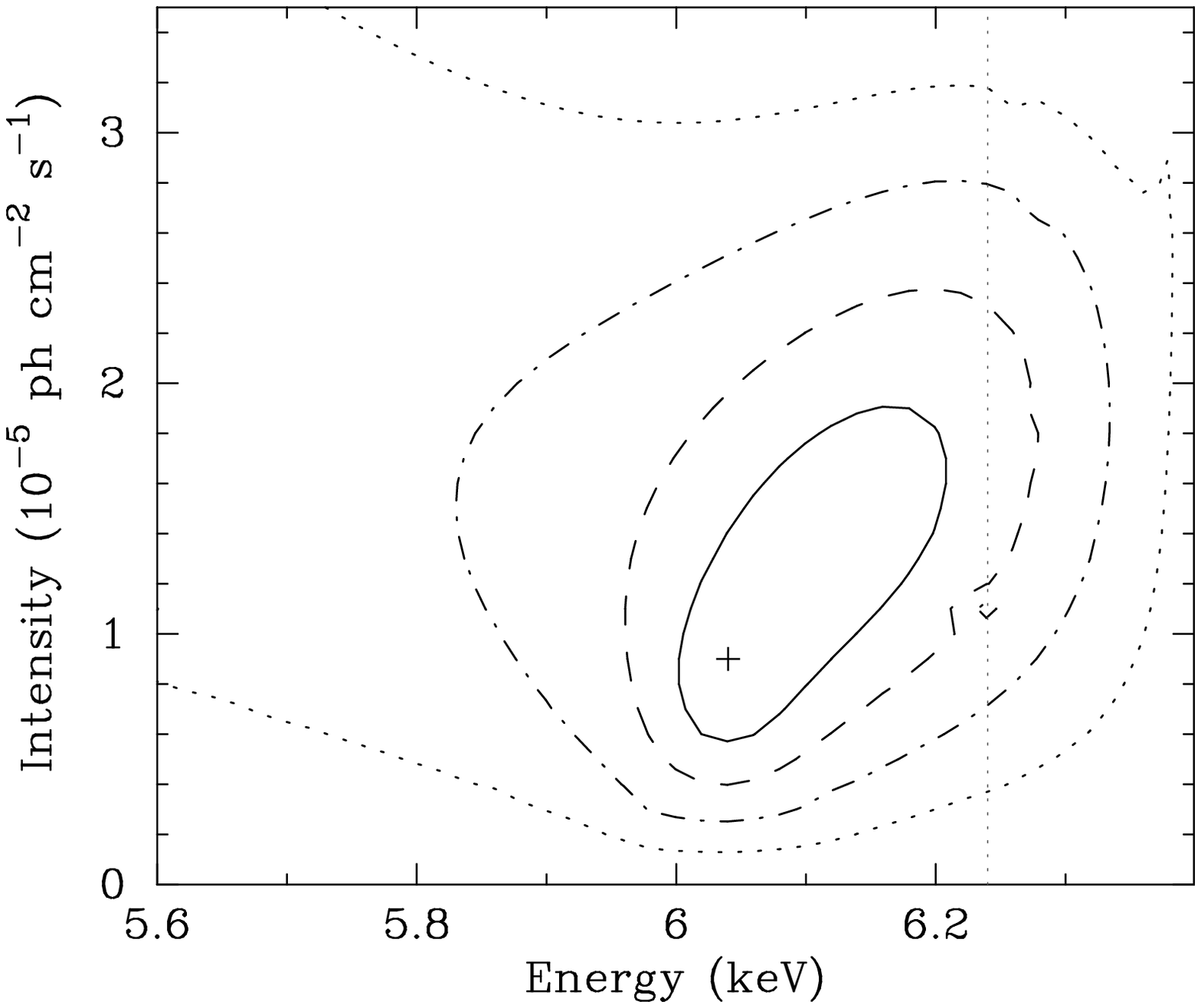}
\caption{Contour plot of intensity versus line centroid energy
for the Gaussian used to model the narrow 6.0 keV emission feature in the pn spectrum.
The width $\sigma$ was left as a free parameter. 1-, 2-, 3- and 4-$\sigma$
confidence levels for two interesting parameters are denoted by solid, dashed, dot-dashed, and dotted lines,
respectively.}
\end{figure}

\begin{figure}
\epsscale{0.75}
\plotone{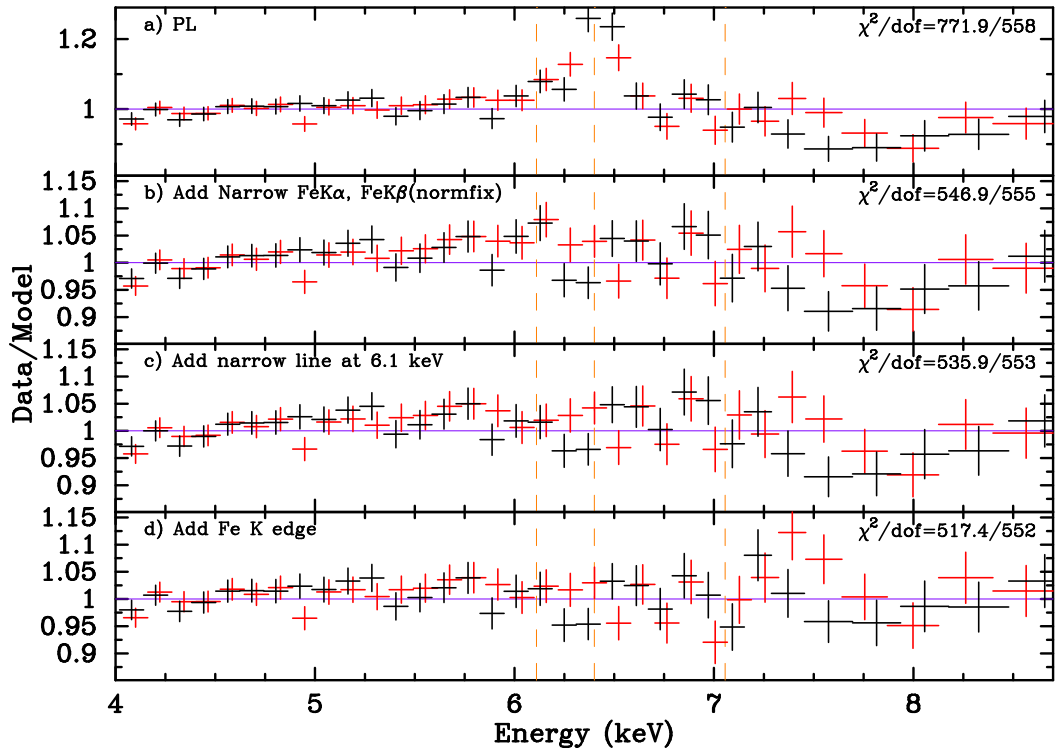}
\caption{Data/model residuals to various model fits to the MOS 1+2 spectrum.
Data are rebinned by a factor of 8. 
Panel {\it a)} shows residuals to a simple power-law model.
In panel {\it b)}, the Fe K$\alpha$ and Fe K$\beta$ emission lines have been
modeled. In panels {\it c)} and {\it d)}, respectively, the 6.11 keV emission
line and the Fe K edge have been added.}
\end{figure}

\begin{figure}
\epsscale{0.60}
\plotone{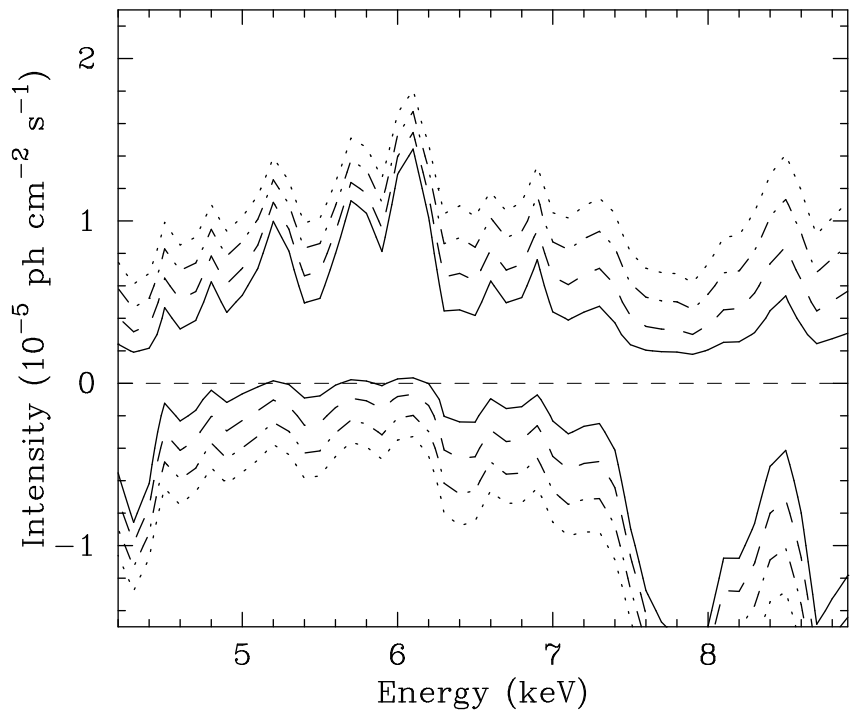}
\caption{Same as Figure 3, but for the MOS 1+2 spectral fitting.}
\end{figure}

\begin{figure}
\epsscale{0.60}
\plotone{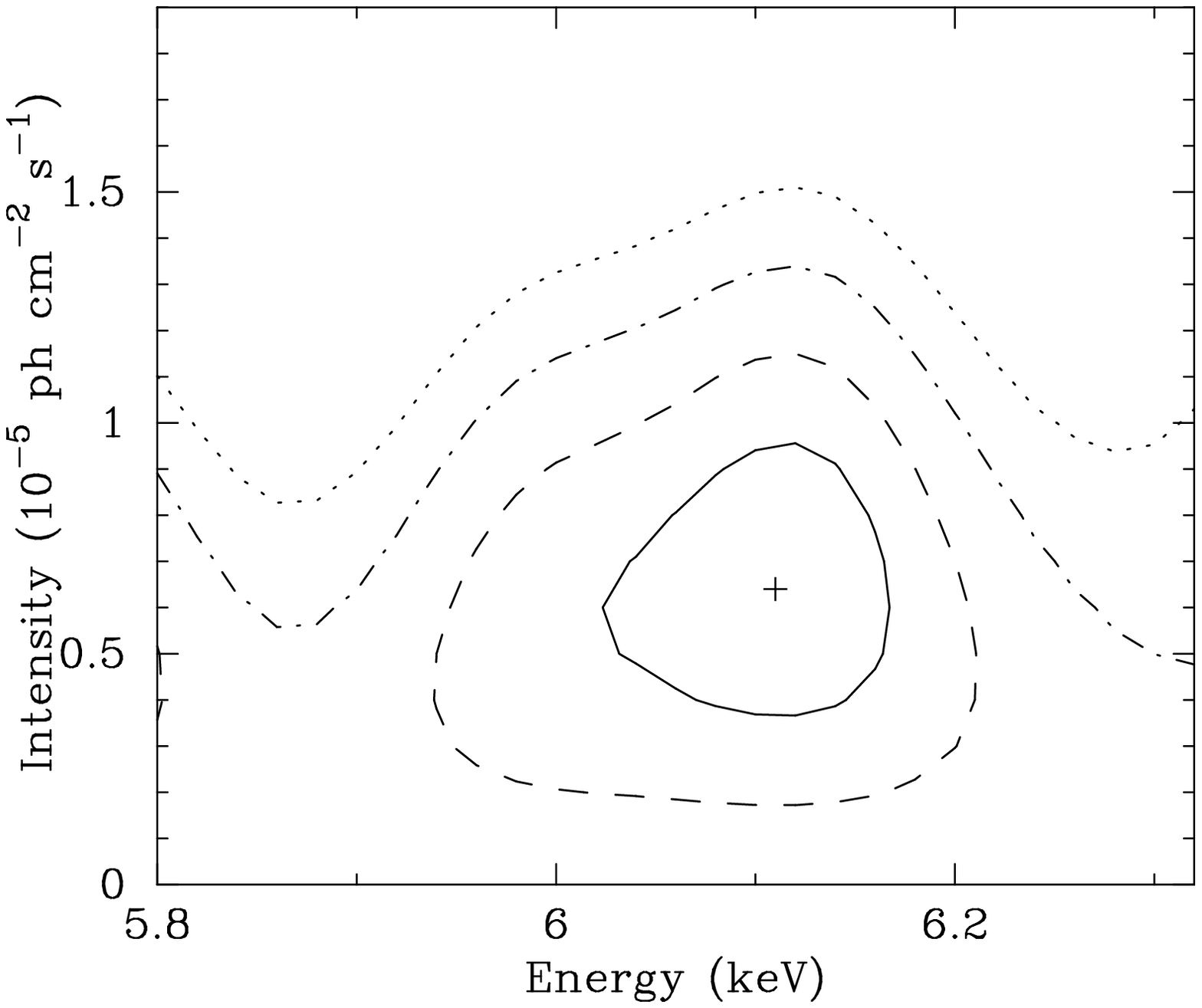}
\caption{Contour plot of intensity versus line centroid energy
for the Gaussian used to model the narrow 6.1 keV emission feature in the MOS 1+2 spectrum.
The width $\sigma$ was left as a free parameter. 1-, 2-, 3- and 4-$\sigma$
confidence levels for two interesting parameters are denoted by solid, dashed, dot-dashed, and dotted lines,
respectively.}
\end{figure}

\begin{figure}
\epsscale{0.65}
\plotone{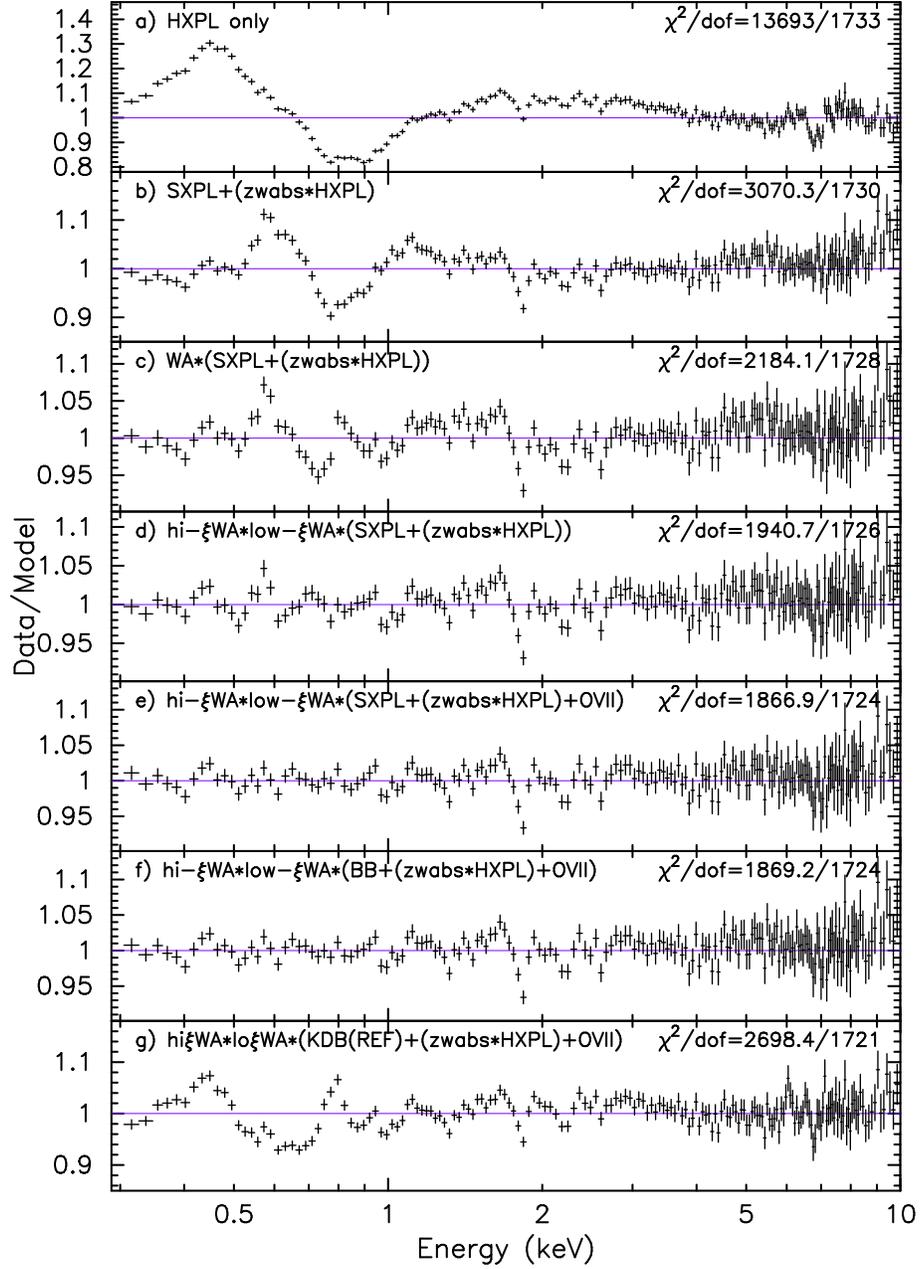}
\caption{Data/model residuals to spectral fits to the 0.2--10 keV
EPIC-pn spectrum. All models include narrow Gaussian components 
to model the Fe K$\alpha$, Fe K$\beta$, Ni K$\alpha$ and 6.04 keV
emission lines, an Fe K edge at 7.11 keV, and a column of neutral gas to model 
Galactic absorption.
Panel {\it a)} shows the results when a simple hard X-ray 
power-law component (HXPL) is used. In Panel {\it b)}, a 
neutral absorbing column has 
been added to the HXPL, and a steep, soft X-ray power-law
component (SXPL) has been added.
In panels {\it c)} and {\it d)}, the low- and hi-ionization
warm absorbers, respectively, have been been included.
In panel {\it e)}, a narrow Gaussian to model \ion{O}{7} emission
has been included; this is the best-fitting ``SXPL'' model.
In panel {\it f)}, the soft power-law has been replaced with a 
blackbody component; this is the best-fitting ``BB'' model.
Panel {\it g)} shows the results from modeling the soft excess 
as blurred reflection from an ionized disk. 
(The small dip near 6.8 keV is not
significant and is likely an artifact of fitting.)}
\end{figure}

\clearpage

\begin{figure}
\epsscale{0.75}
\plotone{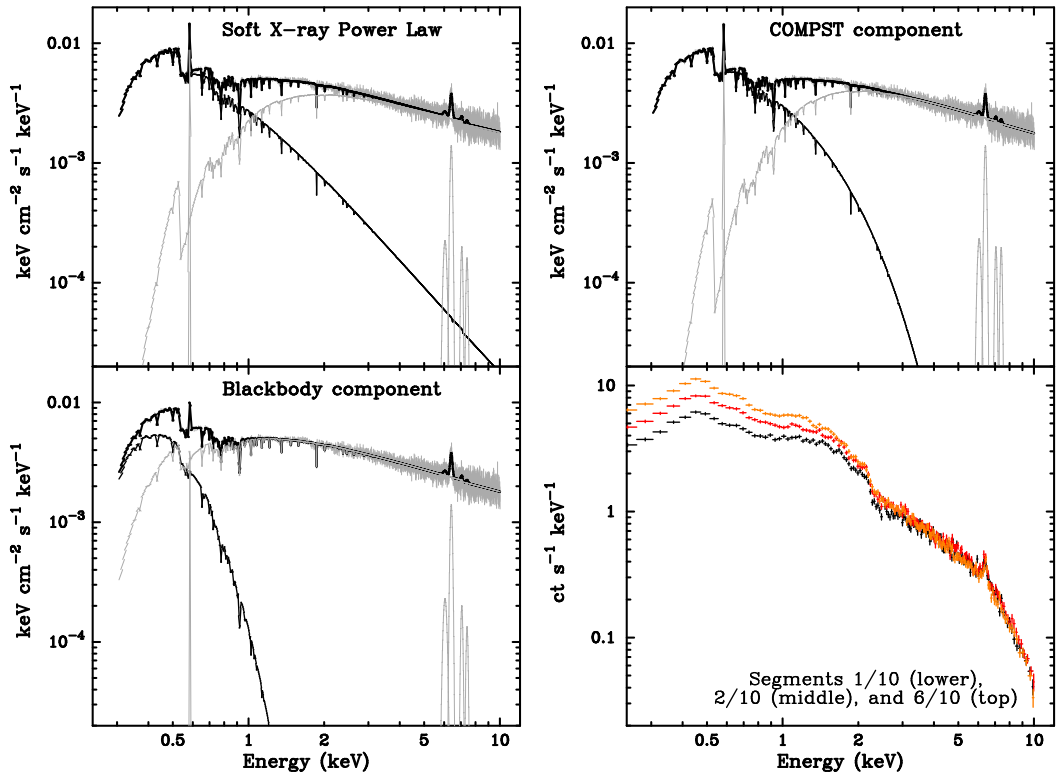}
\caption{Unfolded spectra for the best-fitting SXPL, BB and COMPST models.
Panel {\it d)} shows the spectra for three selected time-resolved
segments, further illustrating the increase in soft X-ray flux
between segments 1, 2 and 6.}
\end{figure}


\begin{figure}
\epsscale{0.75}
\plotone{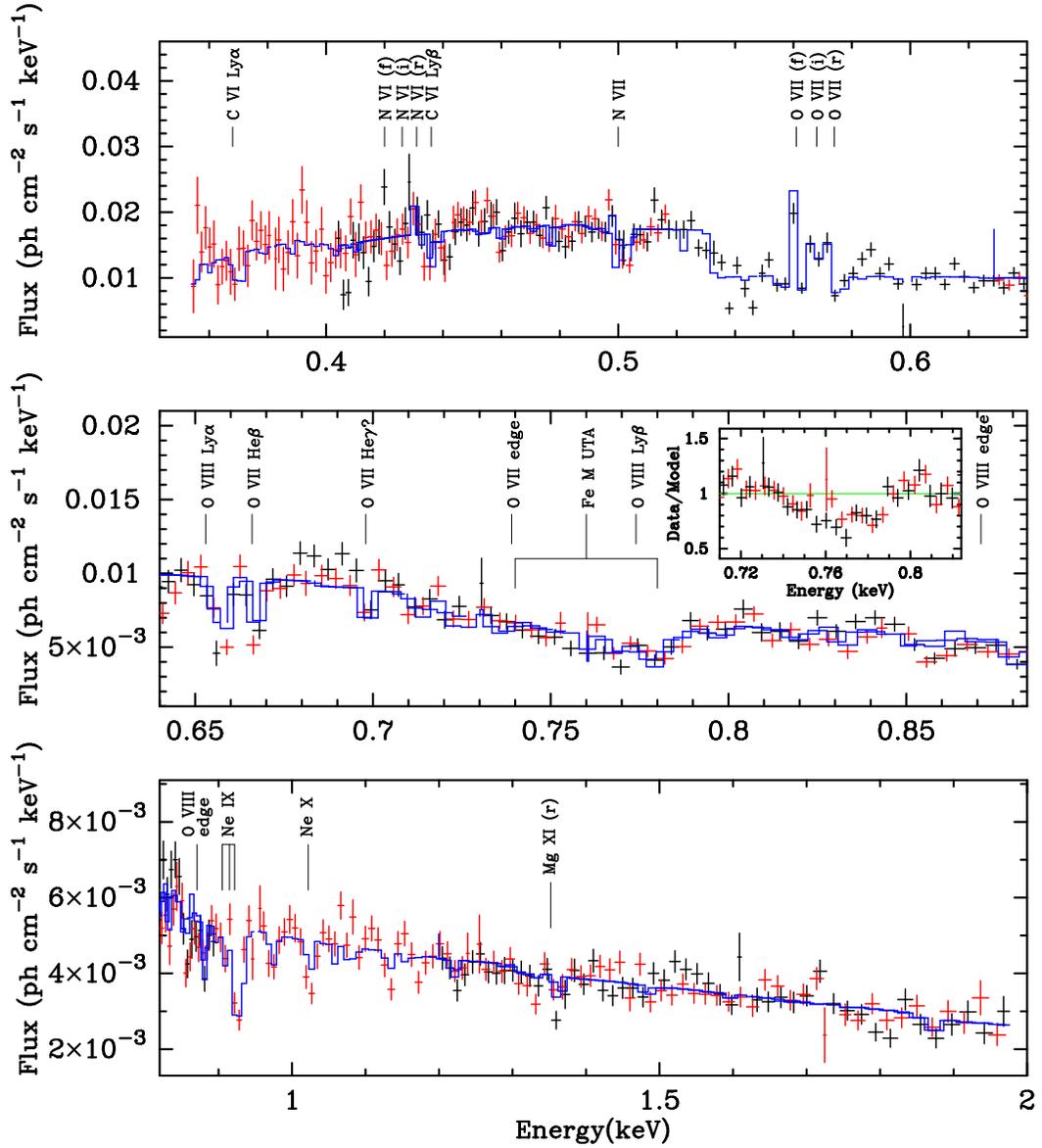}
\caption{The RGS spectrum for NGC 3227.
The solid line shows the best-fitting unfolded model,
described in $\S$5. Data have been rebinned by a factor of 2, and are plotted
in the rest frame. Black and red data points denote RGS 1 and 2, respectively.
The vertical lines, which indicate identified lines and edges, denote
rest-frame (systemic) energies.
The residuals near 687 and 850 eV are 
instrumental artifacts.}
\end{figure}

\begin{figure}
\epsscale{0.60}
\plotone{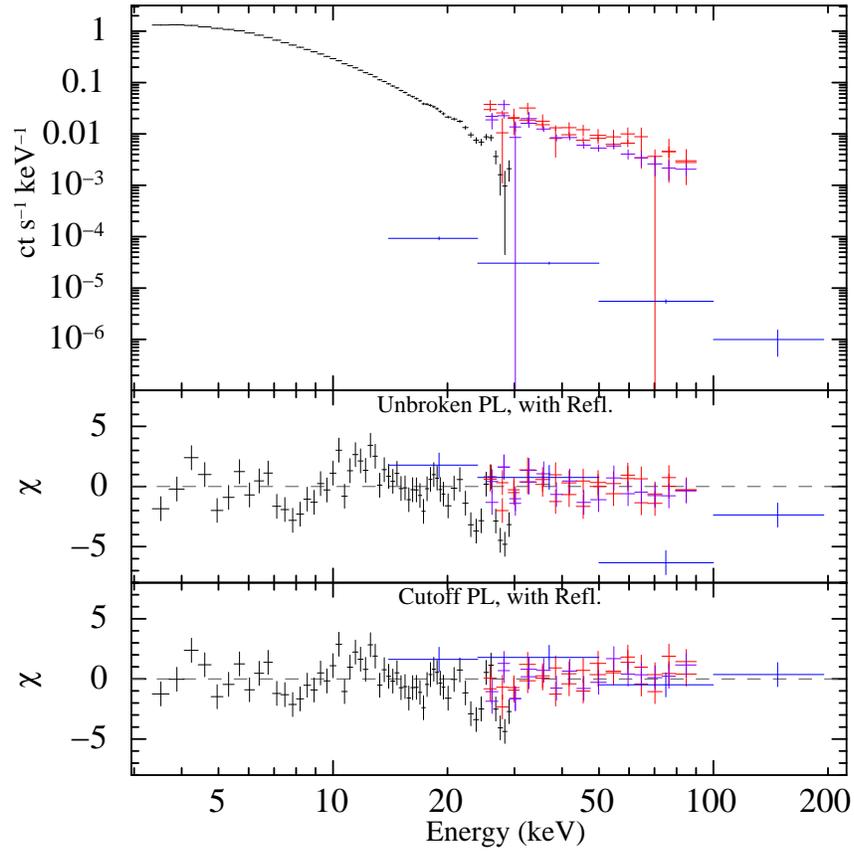}
\caption{The upper panel shows the {\it RXTE}-PCA (black),
{\it RXTE}-HEXTE cluster A (red) and B (purple), and
{\it Swift}-BAT data (blue). The middle and lower panels show 
$\chi$ residuals to the best-fit models with and without a 
high-energy cutoff in the power-law component, respectively. 
In each case, the model
contains a Compton reflection component and Fe K$\alpha$ emission.}
\end{figure}

\begin{figure}
\epsscale{0.40}
\plotone{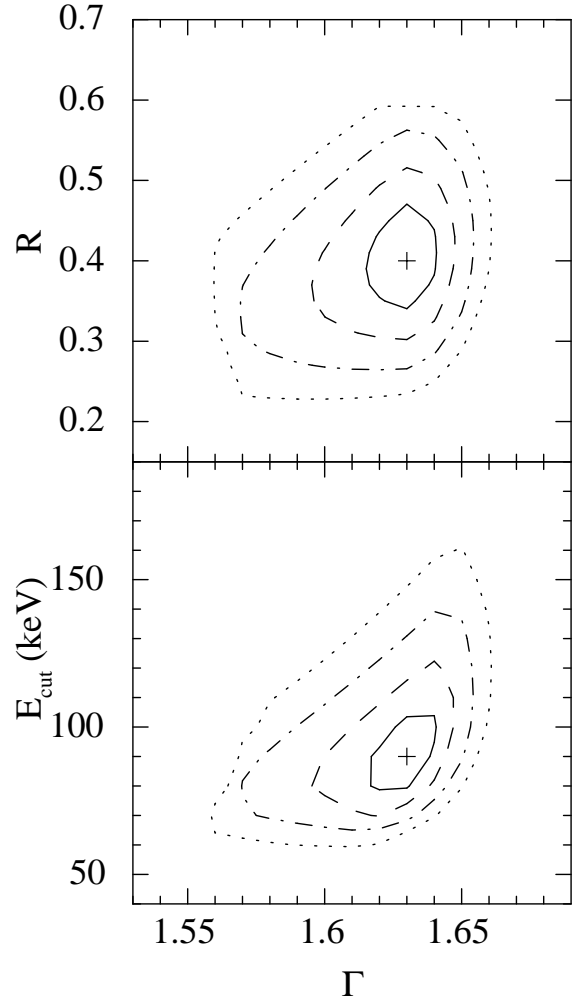}
\caption{Contour plots showing the strength of the Compton reflection
component, $R$ (measured via a {\sc pexrav} component; top panel) 
and cutoff energy (bottom) versus hard X-ray photon index from spectral
fits to {\it RXTE} PCA and HEXTE and {\it Swift}-BAT data.
1-, 2-, 3- and 4-$\sigma$
confidence levels for two interesting parameters
are denoted by solid, dashed, dot-dashed, and dotted lines,
respectively.}
\end{figure}

\begin{figure}
\epsscale{0.50}
\plotone{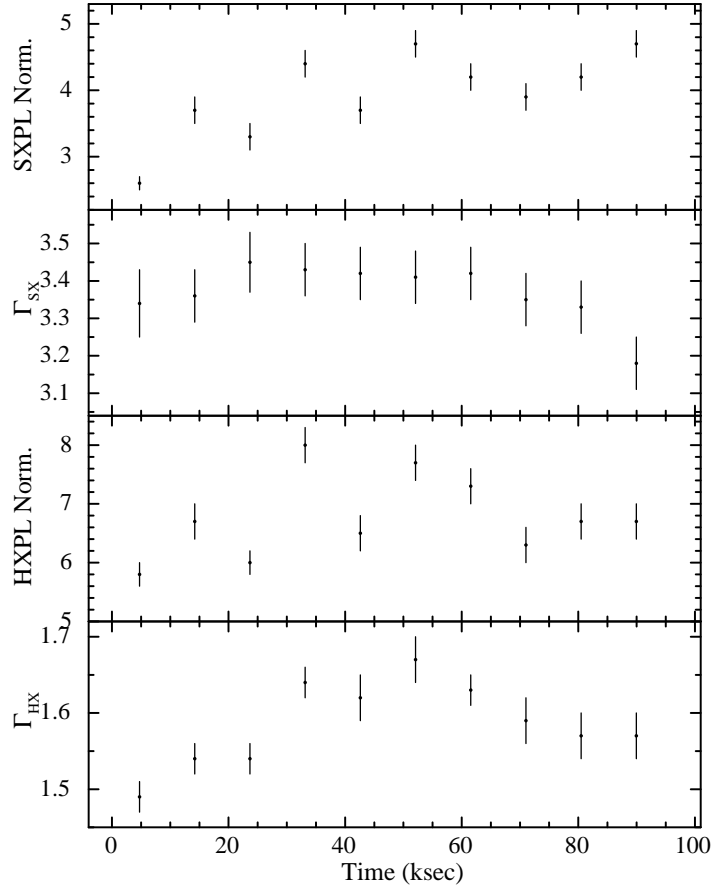}
\caption{The results of fitting the SXPL model to the ten time-resolved spectral segments.
From top to bottom: Normalization of the soft X-ray power-law component,
$\Gamma_{\rm SX}$, normalization of the hard X-ray power-law component, and
$\Gamma_{\rm HX}$.}
\end{figure}

\begin{figure}
\epsscale{0.50}
\plotone{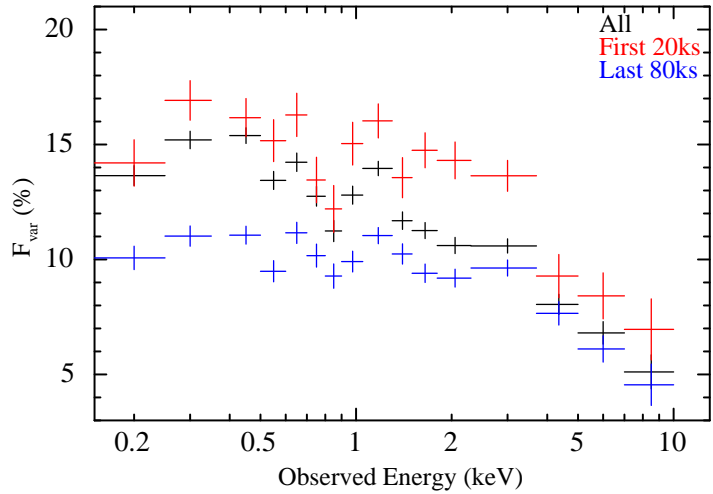}
\caption{$F_{\rm var}$ spectra derived from EPIC-pn data
for the entire duration (black), the first 20 ks (red) and the
final 80 ks (blue).}
\end{figure}

\begin{figure}
\epsscale{0.75}
\plotone{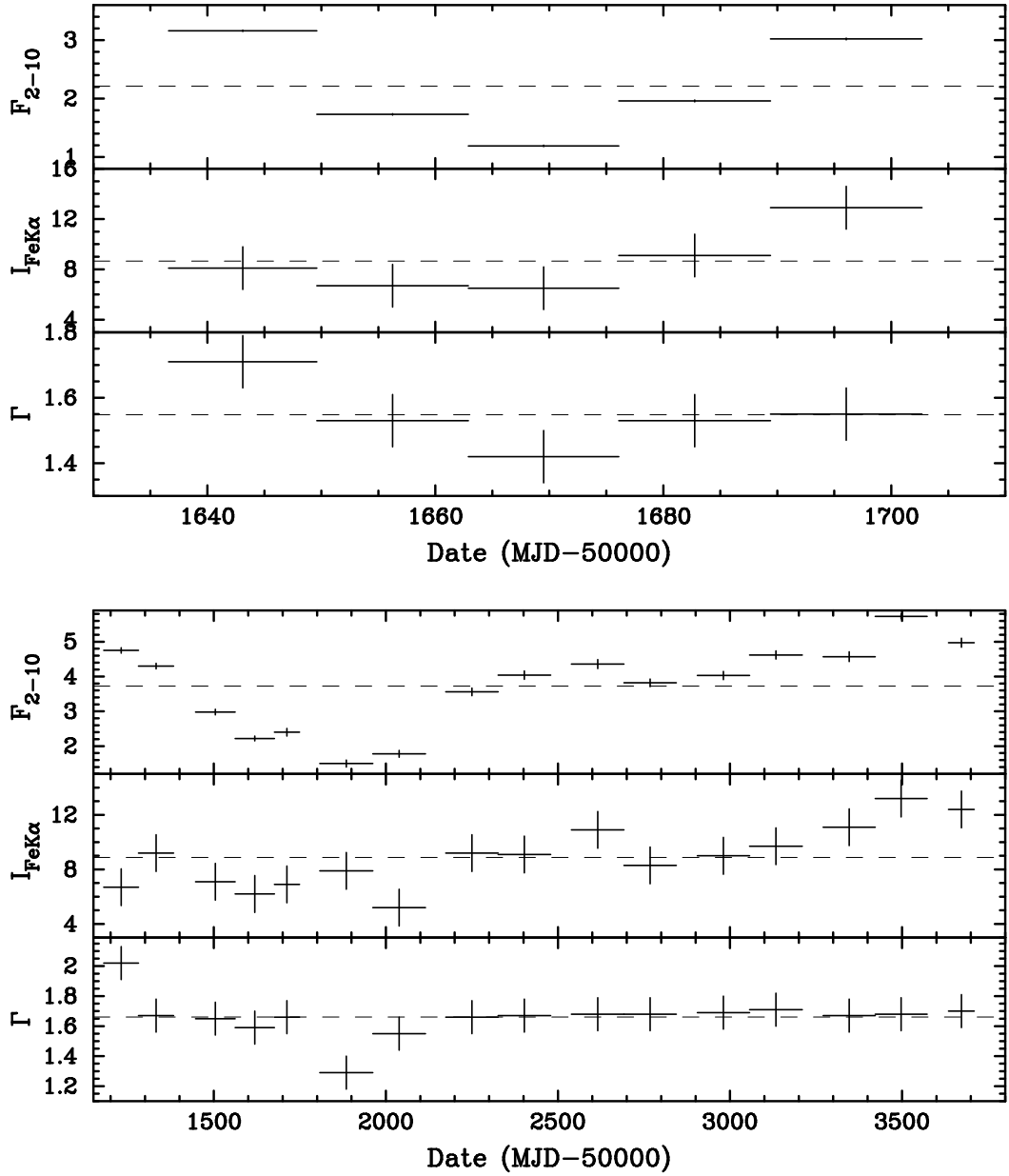}
\caption{The results of time-resolved spectral fits to the medium-term (top) and long-term
(bottom) {\it RXTE} data, showing 2--10 keV continuum flux in units of 10$^{-11}$ erg cm$^{-2}$ s$^{-1}$, 
Fe K line intensity $I_{\rm Fe K\alpha}$ in units of 10$^{-5}$ ph cm$^{-2}$ s$^{-1}$, and
$\Gamma_{\rm HX}$.}
\end{figure}

\begin{figure}
\epsscale{0.40}
\plotone{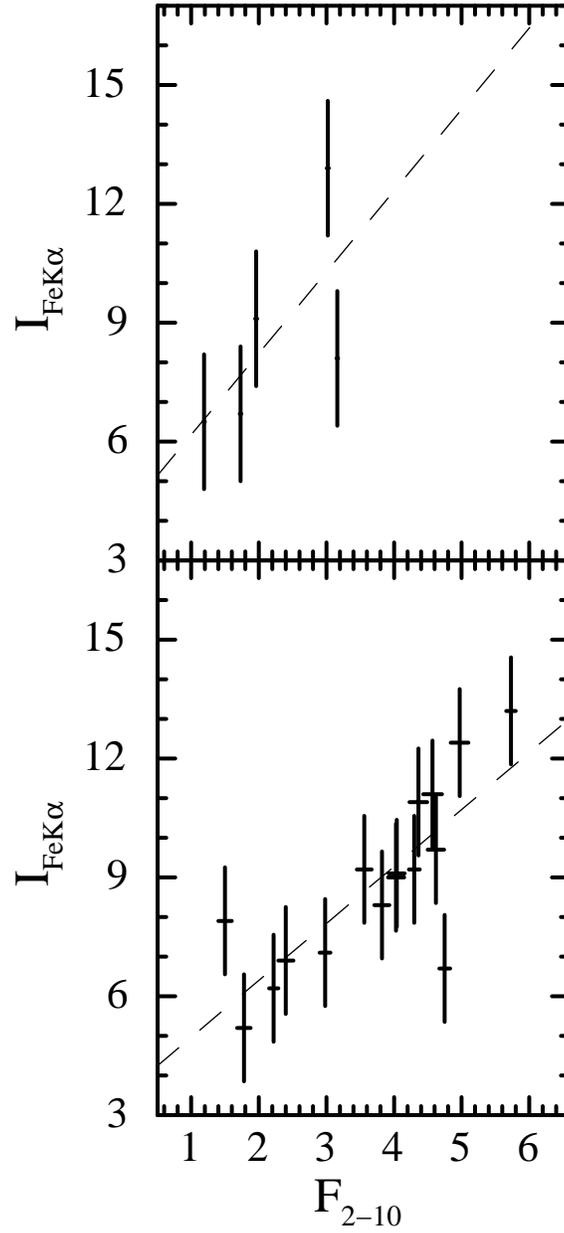}
\caption{Zero-lag correlation diagrams in which Fe line intensity $I_{\rm Fe K\alpha}$,
in units of 10$^{-5}$ ph cm$^{-2}$ s$^{-1}$, is plotted against 
2--10 keV continuum flux in units of 10$^{-11}$ erg cm$^{-2}$ s$^{-1}$. 
Dashed lines indicate the best-fitting linear model.}
\end{figure}


\begin{figure}
\epsscale{0.40}
\plotone{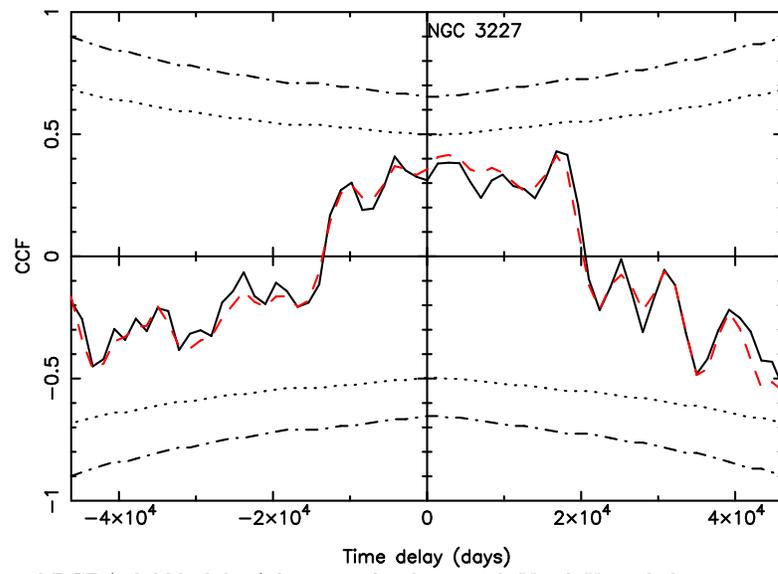}
\caption{ICF (red dashed line) and DCF (solid black line)
functions for the 0.2--1 keV soft X-ray light curve
versus the OM UV continuum light curve (positive lag indicates soft X-ray leading 
the UV.)}
\end{figure}

\end{document}